\documentclass[aps,prb,twocolumn]{revtex4}
\usepackage{bm}
\usepackage{graphicx}

\begin{document}
\title{Electron transport and current fluctuations in short coherent conductors}
\author{Dmitri S. Golubev$^{1,3}$, Artem V. Galaktionov$^{2,3}$, and
Andrei D. Zaikin$^{2,3}$}
\affiliation{$^1$Institut f\"ur Theoretische Festk\"orperphysik,
Universit\"at Karlsruhe, 76128 Karlsruhe, Germany \\
$^{2}$Forschungszentrum Karlsruhe, Institut f\"ur Nanotechnologie,
76021, Karlsruhe, Germany\\
$^{3}$I.E. Tamm Department of Theoretical Physics, P.N.
Lebedev Physics Institute, 119991 Moscow, Russia}

\begin{abstract}
Employing a real time effective action formalism we analyze
electron transport and current fluctuations in comparatively short
coherent conductors in the presence of electron-electron
interactions. We demonstrate that, while Coulomb interaction
tends to suppress electron transport, it may {\it strongly enhance}
shot noise in scatterers with highly transparent conducting channels.
This effect of excess noise is governed by the Coulomb gap observed
in the current-voltage characteristics of such scatterers. We also
analyze the frequency dispersion of higher current cumulants
and emphasize a direct relation between electron-electron interaction
effects and current fluctuations in disordered mesoscopic conductors.

\end{abstract}

\maketitle
\section{Introduction}
Recently considerable progress has been reached in understanding
of an interplay between scattering effects and electron-electron
interactions in low-dimensional disordered conductors. In
particular, a profound relation \cite{GZ00,GGZ02,kn,bn} between
full counting statistics \cite{LLL} (FCS) and electron-electron
interaction effects in mesoscopic conductors has been discovered
and explored. This relation follows directly from the effective
action of the system which can be conveniently derived by
combining the scattering matrix technique with the path integral
description of interactions \cite{Naz,GZ00,GGZ02}.

An important simplification of the formalism amounts to neglecting
the energy independence of scattering matrix. This approximation
is applicable to rather short coherent scatterers with Thouless
energy exceeding all other relevant energy scales. Besides that,
all inelastic processes are assumed to occur in large reservoirs
but not inside the scatterer. For this model an exact expression
for the effective action was derived \cite{GGZ,kn,bn} and studied
under various approximations, such as regular expansion
\cite{GZ00,GGZ02,GGZ} of the action in powers of the ``quantum''
part of the fluctuating field describing interactions as well as
renormalization group (RG) analysis \cite{kn,bn}. Both approaches
are justified in the metallic regime, i.e. provided the effective
system conductance is much larger than the quantum unit $e^2/h$.
Further generalizations of the model allowed to consider spatially
extended metallic conductors and to describe inelastic processes
inside the system \cite{GZ03}.

The main goal of the present paper is to investigate the problem
employing a different set of approximations in order to go beyond
the regimes already studied in the literature. Having in mind the
relation between FCS and interaction effects in mesoscopic
conductors, it would be useful to develop a straightforward and
handy formalism enabling one to analyze the frequency dependence
of higher order current correlation functions. In other words, the
task at hand is to generalize the FCS-type of approach \cite{LLL} (which
is valid only in the low frequency limit) to arbitrary
frequencies. In addition, it is desirable to extent the
description of interaction effects in mesoscopic coherent
conductors beyond the metallic regime. This step would allow, for
instance, for a more detailed analysis of electron-electron
interactions effects in conductors with few conducting channels,
such as, e.g., single-wall carbon nanotubes and organic
molecules.

Here we will demonstrate that the above goals can be accomplished
assuming transmissions $T_n$ of conducting channels to be either
sufficiently small or, on the contrary, sufficiently close to
unity. In these situations significant simplifications of the
exact effective action can be worked out making the whole formalism
more convenient for practical calculations.

Perhaps the most striking result of our analysis concerns the
effect of electron-electron interactions on the shot noise in
conductors with several highly transmitting channels. We will
demonstrate that in this case Coulomb interaction yields {\it
strong enhancement} of the shot noise. Specifically, at
sufficiently small voltages $V$ the low frequency noise power
spectrum ${\cal S}_I(0)$ in weakly reflecting scatterers takes the
form
\begin{equation}
{\cal S}_I(0) \propto |V|^{1-\frac{2}{g}},
\label{enh}
\end{equation} where $g$ is the dimensionless
conductance of a scatterer. Provided $g$ is not much bigger than 2
the result (\ref{enh}) strongly exceeds the non-interacting
dependence \cite{Khlus} ${\cal S}_I(0) \propto |V|$. In the limit
of large voltages the noise spectrum is offset by the value of the
Coulomb gap,
\begin{equation}
{\cal S}_I(0) \propto V +e/2C,
\end{equation}
where $C$ is the effective capacitance of the scatterer. We also
note that interaction-induced excess noise in weakly reflecting
scatterers is observed only at sufficiently low frequencies
$\omega \lesssim eV$, while at larger $\omega$ Coulomb interaction
-- on the contrary -- suppresses the shot noise.

The structure of the paper is as
follows. In Sec. II we will briefly summarize the main steps of
our derivation of the exact effective action for the conductor
described by an arbitrary energy independent scattering matrix.
This general expression will be analyzed in Sec. III in various
limits. The limiting forms of the effective action will then be
used to describe electron transport and current fluctuations in
coherent conductors. In Sec. IV we will use them to obtain the
frequency and voltage dependence of current cumulants. In Sec. V
we will consider a scatterer shunted by a linear Ohmic conductor
and, assuming that the system conductance is sufficiently large,
discuss perturbative logarithmic interaction corrections to the
$I-V$ curve as well as their RG analysis. In Sec. VI and VII we
will go beyond the metallic limit and study the effect of
electron-electron interactions respectively on the $I-V$ curve and
on the shot noise of highly transmitting conductors. Further
extensions of our results for the case of quantum dots will be
considered in Sec. VIII. Some formal manipulations and further
details of our derivations are presented in Appendices A, B and C.

\section{Effective action}

For our derivation we will make use of the Keldysh path integral formalism
describing the system of interacting electrons by the action
which depends on the Grassmann electron fields
$\Psi_\sigma^{1,2}(t,\bm{r}),\Psi^{1,2\,\dagger}_\sigma(t,\bm{r}).$
The labels 1 and 2 correspond to the forward and the backward branches
of the Keldysh contour and $\sigma$ is the spin index.
Performing the standard Hubbard-Stratonovich decoupling of the Coulomb
interaction terms in the action, one introduces the two fluctuating electric
potentials $V_1(t,\bm{r})$  and $V_2(t,\bm{r})$.
Afterwards the electron
fields are integrated out and one arrives at a formally exact effective
action
\begin{eqnarray}
iS=2{\rm Tr}\ln\check G^{-1}_V+i\int_0^t dt'\int d^3\bm{r}
\frac{(\nabla V_1)^2-(\nabla V_2)^2}{8\pi}.
\label{action1}
\end{eqnarray}
Here we assumed the spin degeneracy and defined
\begin{eqnarray}
\check G_V^{-1}=\left(\begin{array}{c} i\frac{\partial }{\partial
t}+\frac{\nabla^2}{2m}+ \mu-U(\bm{r})+eV_1 \hspace{0.9cm} 0
\hspace{0.9cm}\\ \hspace{0.9cm}0 \hspace{0.9cm} i\frac{\partial }{\partial
t}+\frac{\nabla^2}{2m}+\mu-U(\bm{r})+eV_2
\end{array}\right),
\label{GV-1}
\end{eqnarray}
where $U(\bm{r})$ is the external static potential.

The action (\ref{action1}) can also be expressed in a different way, which is convenient for
generating correlation functions of the current:
\begin{eqnarray}
iS=\ln
{\rm Tr}\left[{\cal T}{\rm e}^{-i\int_0^t dt'{\bm H}_1(t')}
\hat{\bm \rho}_0 \tilde {\cal T}{\rm e}^{i\int_0^t dt'{\bm H}_2(t')}\right]
\nonumber\\
+i\int_0^t dt'\int d^3\bm{r}
\frac{(\nabla V_1)^2-(\nabla V_2)^2}{8\pi},
\label{action0}
\end{eqnarray}
with the trace taken over the fermionic variables. Here $\hat{\bm \rho}_0$ is
the initial $N-$particle density matrix of electrons,
\begin{eqnarray}
{\bm H}_{1,2}=\sum_{\sigma}\int d^3{\bm r}\,\hat\Psi^\dagger_\sigma({\bm r})
\hat H_{1,2}(t)\hat\Psi_\sigma({\bm r}),
\nonumber
\\
\hat H_{1,2}(t)=-\frac{\nabla^2}{2m}+U(\bm{r})-eV_{1,2}(t,\bm{r})
\label{H}
\end{eqnarray}
are the effective Hamiltonians on the forward and backward parts of the
Keldysh contour,  ${\cal T}$ and
$\tilde{\cal T}$ are respectively the forward and backward time ordering
operators.

One can also show (see Appendix A for details) that the action (\ref{action1})
can be expressed in the following equivalent form \cite{GGZ}
\begin{eqnarray}
iS&=& 2{\rm tr}\ln\left\{1+( {\cal U}_2(0,t) {\cal U}_1(t,0)-1)\hat\rho_0\right\}
\nonumber\\ &&
+\,i\int_0^t dt'\int d^3\bm{r}\frac{(\nabla V_1)^2-(\nabla V_2)^2}{8\pi},
\label{action2}
\end{eqnarray}
where
\begin{eqnarray}
 {\cal U}_{1,2}(t_1,t_2)={\cal T}e^{-i\int_{t_1}^{t_2}dt'\left(
 -\frac{\nabla^2}{2m}-
\mu+U(\bm{r})+eV_{1,2}(t',\bm{r})\right)} \label{U12}
\end{eqnarray}
are the evolution operators. Here we distinguish between the ``full''
trace  (``${\rm Tr}$'', Eq. (\ref{action1})) implying
the integration over both the coordinates and the time variable, and
the ``reduced'' trace  (``${\rm tr}$'',
Eq. (\ref{action2})) which denotes the coordinate integration only.

In this paper we will analyze the properties of a short coherent
conductor placed in-between two bulk reservoirs. Electron transport
across such a conductor can be described by the scattering matrix:
\begin{equation}
\hat S=\left(\begin{array}{cc} \hat r & \hat t' \\ \hat t & \hat r'
\end{array}
\right),
\label{scm}
\end{equation}
Below we will assume that this matrix does not depend on the
energy of incoming electrons. This assumption is justified provided the dwell
time of an electron inside the scatterer is shorter than any other
relevant time scale. In addition, we assume that an external circuit,
which also includes the leads, is linear and
characterized by an effective impedance $Z_S(\omega).$
The fluctuating electric potential is assumed to slowly depend on
the coordinates in the reservoirs, but to drop sharply across the scatterer.
Denoting this voltage drop as $V_{1,2}(t')$, one can also introduce the
corresponding Keldysh phase fields
$\varphi_{1,2}(t)=\int_0^t dt'eV_{1,2}(t')$ as well as their combinations
$\varphi^+=(\varphi_1+\varphi_2)/2$ and $\varphi^-=\varphi_1-\varphi_2$.
The total action of our system reads
\begin{eqnarray}
iS[\varphi^\pm]=iS_S[\varphi^\pm]+i\int_0^tdt'\,
C\frac{\dot\varphi^+\dot\varphi^-}{e^2}+ iS_{\rm sc}[\varphi^\pm]
. \label{S}
\end{eqnarray}
Here $iS_S[\varphi^\pm]$ describes the external
circuit:
\begin{eqnarray}
iS_S[\varphi^\pm]= -\frac{1}{2e^2}\int_0^t
dt_1dt_2\,\varphi^-(t_1)\alpha_S(t_1-t_2)\varphi^-(t_2)
\nonumber\\
+\,\frac{i}{e^2}\int_0^t
dt_1dt_2\,\varphi^-(t_1)Z^{-1}_S(t_1-t_2)\big(eV_x-\dot\varphi^+(t_2)\big), \label{SS}
\end{eqnarray}
where $Z^{-1}_S(t)=\int\frac{d\omega}{2\pi}\,\frac{{\rm e}^{-i\omega
t}}{Z_S(\omega)}$ is the response function of the external shunt,
$\alpha_S(t)=\int\frac{d\omega}{2\pi}\,{\rm e}^{-i\omega t}{\rm
Re}\left(\frac{\omega\coth\frac{\omega}{2T}}{Z_S(\omega)}\right),$
and $V_x$ is the total voltage drop on the system ``scatterer $+$ external shunt". 

The second term in Eq. (\ref{S}) originates from the last term of
Eq. (\ref{action2}) and describes the energy of the fluctuating fields.
This term  contains the scatterer capacitance $C.$

Finally, the most interesting for us contribution $iS_{\rm sc}$ is given
by the first term of Eq. (\ref{action2}) with the evolution operators
${\cal U}_{1,2}$ describing the scattering of electrons in the conductor.
Following the standard procedure we introduce the scattering channels
(labeled by indices) and denote the coordinate and the group velocity in
the $k-$th channel respectively as $y_k$ and $v_k.$ The matrix
corresponding to the operator ${\cal  U}_1(t,0)$ then takes the form
\begin{eqnarray}
{\cal
U}^{nk}_1(t,0;y_2,y_1)=e^{i\alpha_n\varphi_{1}(t)-i\alpha_k\varphi_1(0)}
\frac{\delta\left(\frac{y_2}{v_n}-\frac{y_1}{v_k}-t\right)}{\sqrt{v_nv_k}}\times
 \nonumber\\
\big[\delta_{nk}+\theta(y_2)\theta(-y_1)e^{-i\alpha_n\varphi_1(-y_1)
+i\alpha_k\varphi_1(-y_1)} (S_{nk}-\delta_{nk})\big],\nonumber\\ \label{U}
\end{eqnarray}
where $S_{nk}$ is the matrix element of the $S-$matrix (\ref{scm}),
$\alpha_k=1$ provided the $k-$th channel is in the left lead and
$\alpha_k=0$ otherwise. The operator ${\cal U}_2(0,t)$ is constructed from
${\cal U}_1(t,0)$ by means of the Hermitian conjugation together with the
replacement $\varphi_1\to\varphi_2$. We point out that Eq. (\ref{U}) is
similar to the well known relation between the Green function  and the
$S-$matrix established by Fisher and Lee \cite{FL}.

Having found the expressions for the evolution operators, we make use of
Eq. (\ref{action2}) and arrive at the final result:
\begin{widetext}
\begin{equation}
iS_{\rm sc}=2{\rm tr}\ln\left\{\hat 1\delta(x-y) +\theta(t-x)\theta(x)
\left[\begin{array}{cc} \hat t^\dagger\hat t({\rm e}^{i\varphi^-(x)}-1) &
2i \hat t^\dagger\hat r' \sin\frac{\varphi^-(x)}{2} \\  2i \hat r^{\prime
\dagger}\hat t \sin\frac{\varphi^-(x)}{2} & \hat t^{\prime \dagger}\hat
t'( {\rm e}^{ -i\varphi^-(x)}-1)
\end{array}
\right] \left[\begin{array}{c} \rho_0(y-x){\rm
e}^{i\frac{\varphi^+(x)-\varphi^+(y)}{2}}\hspace{0.9cm}  0
\\  0\hspace{0.9cm}  \rho_0(y-x){\rm e}^{i\frac{\varphi^+(y)-\varphi^+(x)}{2}}
\end{array}
\right] \right\}, \label{S1}
\end{equation}
\end{widetext}
where
\begin{equation}
\rho_0(x)=\int \frac{dx}{2\pi} \frac{{\rm e}^{iEx}}{1+{\rm
e}^{E/T}}=\frac{1}{2}\delta(x)-\frac{iT}{2\sinh\pi Tx} \label{rho0}
\end{equation}
is the Fourier
transform of the Fermi function.

\section{Limiting forms of the action}

The general expression for the effective action (\ref{S1})
contains a great deal of information and can be used in order to
describe a variety of different phenomena. At the same time this
expression still remains rather complex, and further simplifications
are highly desirable. These simplifications can be achieved in various
limiting cases to be discussed below in this section.

Before we turn to concrete calculations let us note that several important
physical limits of Eq. (\ref{S1}) are already well known and have been
studied in
details. For instance, setting  $\varphi^-={\rm const}$ and $\dot\varphi^+=eV$
one reduces the action (\ref{S1}) to the FCS cumulant generating
function
\cite{LLL} for a coherent scatterer in the absence of electron-electron
interactions. In this case $\varphi^-$ plays the role of the so-called counting field.

If one allows for the time-dependent voltage $V(t)$ across the scatterer
and at the same time keeps $\varphi^-$ constant,
with the aid of Eq. (\ref{S1}) one can describe FCS at an ac bias,
adiabatic pumping through the conductor and related effects \cite{Lev1}.
In addition, neglecting the interactions but keeping the full time dependence
for the phase  $\varphi^-$, with the aid of Eq. (\ref{S1})
one can fully describe frequency dispersion of all current cumulants.
For this purpose it suffices to perform a regular expansion of the
action (\ref{S1}) in powers of the field $\varphi^-$. Frequency dispersion
of the third current cumulant was analyzed in this way in Ref.
\onlinecite{GGZ}.

The same expansion allows to obtain valuable information about the effect
of electron-electron interactions on transport properties, shot noise as
well as higher cumulants of the current operator. Assuming the
dimensionless conductance of a scatterer $g$ and/or that of the external
circuit $g_S$ are/is large, one can obtain the interaction correction to
the current\cite{GZ00} of order $1/(g+g_S)$ expanding the action
(\ref{S1}) up to the second order in $\varphi^-$. In order to derive the
interaction correction to the shot noise \cite{GGZ02} one should expand
the action up to the third order in  $\varphi^-$.

Finally, keeping the exact non-linear dependence of the action on the
fields $\varphi^\pm$ but expanding (\ref{S1}) to the first order in the
transmission matrix $\hat t^\dagger\hat t$, one immediately reproduces the
well known AES effective action \cite{ESA,SZ}. The latter action, being
combined with $iS_{S}$ (\ref{SS}) and the capacitance term, describes
Coulomb blockade effects in tunnel junctions embedded in a linear
electromagnetic environment\cite{SZ,IN}.

In what follows we will investigate the properties of the action (\ref{S1})
in some other limiting cases.

\subsection{Weakly transmitting barriers}

Let us first consider the case of weakly transmitting barriers and expand
the action in powers of the matrix $\hat t^\dagger\hat t$. As we have
already pointed out, the first order terms of this expansion yield the AES
effective action \cite{ESA,SZ}. Here we proceed further and expand the
action (\ref{S1}) up to $(\hat t^\dagger\hat t)^2$ keeping the complete
nonlinear dependence on the fluctuating phases $\varphi^\pm$. It is easy
to see that in order to recover all such terms it is necessary to expand
the logarithm in Eq. (\ref{S1}) up to the fourth order in the term
containing the density matrix $\rho_0$. Higher order terms of this
expansion can be omitted within the required accuracy since they do not
contain contributions proportional to $T_n$ and $T_n^2$, where $T_n$ represent
the channel transmissions defined in a standard manner as the eigenvalues
of the matrix $\hat t^\dagger\hat t$.

The whole calculation is performed in a straightforward manner, although
requires some care. The final result reads
\begin{widetext}
\begin{eqnarray}
 iS&=&-\frac{i}{\pi}\int_0^{t} dx\;
\dot\varphi^+(x)\sin\varphi^-(x)
\left[{\rm tr}[\hat t^\dagger\hat t]+\frac{2}{3}{\rm tr}[(\hat t^\dagger\hat
t)^2]\sin^2\frac{\varphi^-(x)}{2}\right]
\nonumber\\&&
-\frac{2}{\pi}\int_{0}^{t}dxdy\;\alpha(x-y)\sin\frac{\varphi^-(x)}{2}
\sin\frac{\varphi^-(y)}{2} \left\{ {\rm tr}[\hat t^\dagger\hat t(1-
\hat t^\dagger\hat t)]\cos[\varphi^+(x)-\varphi^+(y)]
+{\rm tr}[(\hat t^\dagger\hat t)^2]\cos\frac{\varphi^-(x)-\varphi^-(y)}{2}
 \right\}
\nonumber\\&&
+\frac{4i}{3}{\rm tr}[(\hat t^+\hat t)^2]\int_0^{t} dx  dy dz\;
\frac{T^3\sin\frac{\varphi^-(x)}{2} \sin\frac{\varphi^-(y)}{2}
\sin\frac{\varphi^-(z)}{2}}{\sinh[\pi T(y-x)]
\sinh[\pi T(x-z)]\sinh[\pi T(z-y)]}
\nonumber \\
&&\times \left\{
\sin[\varphi^+(y)-\varphi^+(x)]\cos\frac{\varphi^-(z)}{2}+
\sin[\varphi^+(z)-\varphi^+(y)]\cos\frac{\varphi^-(x)}{2}+
\sin[\varphi^+(x)-\varphi^+(z)]\cos\frac{\varphi^-(y)}{2}\right\}
\nonumber\\ &&
-16{\rm tr}[(\hat t^\dagger\hat t)^2] \int_0^{t} dx dy dz dw\;
 \rho_0(y-x)\rho_0^*(x-w)\rho_0(w-z)\rho^*_0(z-y)
\nonumber\\ && \times
\sin\frac{\varphi^-(x)}{2} \sin\frac{\varphi^-(y)}{2}
\sin\frac{\varphi^-(z)}{2}\sin\frac{\varphi^-(w)}{2}\cos\left[
\varphi^+(x)-\varphi^+(y)+\varphi^+(z)-\varphi^+(w) \right].
\label{aes2}
\end{eqnarray}
\end{widetext}
where
\begin{eqnarray}
\alpha(x)=\int\frac{d\omega}{2\pi} \,e^{-i\omega x}\omega\coth\frac{\omega}{2T}
=-\frac{\pi T^2}{\sinh^2\pi Tx}.
\end{eqnarray}
Eq. (\ref{aes2}) represents the complete expression
for the effective action valid up to the second order in the
transmissions $T_n$. This expression involves no further
approximations and fully accounts for the non-linear dependence on
the fluctuating phase fields $\varphi^\pm$.

\subsection{Reflectionless limit}

Another physically important limit is that of reflectionless barriers
$\hat r=0$. In this limit the action (\ref{S1}) can be
significantly simplified by means of the exact procedure which we outline
below.

To begin with, we notice that in the case of reflectionless barriers
the action (\ref{S1}) reveals significant similarity
to the Luttinger model of 1D interacting electron gas. The latter
model is usually treated by means of the bosonization technique.
In the case of quantum dots and point contacts this method
was applied in Refs. \onlinecite{Flens,Matveev,ABG,Br}.
One can also evaluate an effective action for reflectionless barriers directly
without employing the bosonization technique. Actually an important feature
of the final result can be guessed even before doing the calculation.
Indeed, since the RPA approximation is known \cite{DL} as
an exact procedure for the Luttinger model,
one can expect that in the limit $\hat r'\to 0$ the action (\ref{S1})
should become quadratic in $\varphi^\pm$,
at least for not very large values of  the phase.
We will demonstrate that this is indeed the case if $|\varphi^-|<\pi$.

Let us put $\hat r'=0$ in Eq. (\ref{S1}). Then the action can be
split into two parts
\begin{eqnarray}
iS^{(0)}_{sc}= iN_{\rm ch}S_0\big[\theta(t-x)\theta(x)({\rm
e}^{i\varphi^-(x)}-1),\;-\varphi^+\big]
\nonumber\\
+\,iN_{\rm
ch}S_0\big[\theta(t-x)\theta(x)({\rm e}^{-i\varphi^-(x)}-1),\;\varphi^+\big],
\label{Ssc0}
\end{eqnarray}
where $N_{\rm ch}$ is the number of open channels and
\begin{equation}
iS_0[a,\varphi^+]=2{\rm tr}\ln\left[1  +
a(x)\rho_0(y-x){\rm e}^{i\frac{\varphi^+(y)-\varphi^+(x)}{2}}\right].
\label{S0}
\end{equation}

The action (\ref{S0}) can be evaluated
exactly. The details of our derivation are summarized
in Appendix B. Here we only present the final result:
\begin{eqnarray}
iS_0[a,\varphi^+]= 2\int
dx\left(\rho_0(0)+\frac{\dot\varphi^+(x)}{4\pi}\right)\ln[1+a(x)]
\nonumber\\
 +\,\int dxdy\, \alpha(x-y)\frac{\ln[1+a(x)]\ln[1+a(y)]}{4\pi },
\label{s0}
\end{eqnarray}
where $\rho_0(0)$ is a large constant,  which
has a meaning of the electron density, and which is later canceled by the corresponding
contribution of ions.
Combining this formula with Eq. (\ref{Ssc0}) we arrive at the main result
of this subsection
\begin{eqnarray}
&& iS_{\rm sc}^{(0)}=-\frac{iN_{\rm ch}}{\pi}\int_0^t
dx\,W(\varphi^-(x))\dot\varphi^+(x)
\nonumber\\ &&
 -\,\frac{N_{\rm ch}}{2\pi}\int_0^t dxdy\,W(\varphi^-(x))\alpha(x-y)W(\varphi^-(y)),
\label{Ssc1}
\end{eqnarray}
where $W(\varphi)$ is the $2\pi-$periodic function of $\varphi^-$,
which equals to $\varphi^-$ in the interval $-\pi<\varphi^-<\pi$.
Under certain conditions the contribution of large phase values,
$|\varphi^-|>\pi$, can be disregarded, and one can use the Gaussian action
\begin{eqnarray}
iS_{\rm sc}^{(0)}\to -\frac{iN_{\rm ch}}{\pi}\int_0^t
dx\,\varphi^-(x)\dot\varphi^+(x)
\nonumber\\
 -\,\frac{N_{\rm
ch}}{2\pi}\int_0^t dxdy\,\varphi^-(x)\alpha(x-y)\varphi^-(y)
\label{Ssc}
\end{eqnarray}
instead of the exact one (\ref{Ssc1}). Within this approximation the
action for the scatterer with $N_{\rm ch}$ perfectly transmitting channels
coincides with that for a linear Ohmic resistor with the conductance
$e^2N_{\rm ch}/\pi.$ It is important to emphasize, however, that this
approximation ignores the action periodicity in the phase space and,
hence, becomes inadequate as soon as electron charge discreteness turns
out to be an important effect.

\subsection{Weakly reflecting barriers}

Let us now assume that the reflection probabilities are  small,
$R_n=1-T_n\ll 1$, but not equal to zero. In this case it is
convenient to proceed perturbatively expanding the action
(\ref{S1}) in powers of $R_n$ or, which is the same, in powers of the
matrices $\hat r',\hat r^{\prime \dagger}$. The details of our
derivation are provided in Appendix C.
Expanding the action (\ref{S1}) to the first order
in $R_n$, one finds
\begin{equation}
iS_{\rm sc}=iS_{\rm sc}^{(0)}+iS_{\rm sc}^{(1)},
\label{smallR}
\end{equation}
where $iS_{\rm sc}^{(0)}$ is defined in Eq. (\ref{Ssc1}) and
\begin{eqnarray}
iS_{\rm sc}^{(1)}&=&\frac{i{\cal R}}{\pi}\int_0^t
dx\,\dot\varphi^+(x)\sin\varphi^-(x)
\nonumber\\ &&
+\,\frac{\cal R}{\pi}\int dxdy\,\alpha(x-y)\,
\varphi^-(x)\sin\varphi^-(y) \nonumber\\ && -\,\frac{2\cal R}{\pi}\int_0^t
dxdy\,\alpha(x-y)\,  e^{i[\varphi^+(x)-\varphi^+(y)]}
\nonumber\\ &&\times\,
e^{\int_0^t
dz[T\coth\pi T(x-z)-T\coth\pi T(y-z)]\varphi^-(z)}
\nonumber\\ &&
\times\,\sin\frac{\varphi^-(x)}{2}\sin\frac{\varphi^-(y)}{2}.
\label{action}
\end{eqnarray}
As we have already pointed out, this expression is justified
provided all the channels are weakly reflecting, $R_n\ll 1$. At
the same time the parameter ${\cal R}=\sum_nR_n$ needs not to be
small should the total number of channels in the system be large.
We also note that different limiting forms of the action
(\ref{aes2}) and (\ref{Ssc1})-(\ref{action}) can be combined if
some channels are weakly transmitting $T_n \ll 1$ while the others
have small reflection coefficients $R_m \ll 1$.

\section{Frequency dispersion of higher current cumulants}

Let us now make use of the above limiting expressions for the
effective action in order to describe various properties of short
coherent conductors. In this section we neglect the effects of
electron-electron interactions and set $\varphi^+(t)=eVt,$ where $V$ is
the time-independent bias voltage.

Complete information about all correlation functions of the
current operator in the zero frequency limit is contained in the
FCS cumulant generating function \cite{LLL}. This function,
however, becomes insufficient if one is interested in the
frequency dependence of the current cumulants. This dependence can
in general be recovered only from the complete effective action
(\ref{S1}). Unfortunately the latter appears too complicated to directly
proceed with the analysis of the $n$-th current cumulant. The situation
 is significantly simplified in the two complementary limits of small
and almost perfect channel transmissions, respectively $T_n \ll 1$ and $1-T_n
\ll 1$. In these two cases one can make use of the limiting forms of the
effective action derived in the previous section. The corresponding analysis
is presented below.

In the absence of interaction effects one can treat the action $iS_{\rm sc}$
as a generating functional for the current correlation functions
\begin{equation}
\left\langle I(t_1)\dots I(t_m)\right\rangle=\left.\frac{(ie)^m \delta^m\,
e^{iS_{\rm sc}[eVt,\varphi^-]}}{\delta\varphi^-(t_1)\dots
\delta\varphi^-(t_m)} \right|_{\varphi^- =0}. \label{cumul}
\end{equation}
Here $I(t_j)$ are measurable classical currents which commute with
each other \cite{KN}. On a quantum level, the same correlator can
be written in terms of the current operators $\hat I(t_j)$,
however the choice of the time ordering becomes important in this
case.

One can also define the $m$-th current cumulant
$\tilde{\cal S}_{m}(t_1,\dots t_m)$
as an irreducible part of the correlator (\ref{cumul}):
\begin{equation}
\tilde{\cal S}_{m}=\left.(ie)^m\,
\frac{\delta^m\;iS_{\rm sc}[eVt,\varphi^-]}{\delta\varphi^-(t_1)
\dots\delta\varphi^-(t_{m})}\right|_{\varphi^-=0}.
\label{Sn}
\end{equation}
For classical currents, the first three cumulants are defined as
\begin{eqnarray}
&& \tilde{\cal S}_{1}=\left\langle I\right\rangle,\nonumber\\ &&
\tilde{\cal S}_{2}(t_1,t_2)=\left\langle \delta I(t_1)\delta I(t_2)
\right\rangle,\nonumber\\ && \tilde{\cal S}_{3}(t_1,t_2,t_3)=\left\langle
\delta I(t_1)\delta I(t_2)\delta I(t_3) \right\rangle ,
\end{eqnarray}
where $\delta I(t)=I(t)-\left\langle I\right\rangle$. The fourth and higher
cumulants take a more complicated form, e.g.,
\begin{eqnarray}
 && \tilde{\cal S}_{4}(t_1,t_2,t_3,t_4)=\left\langle
\delta I(t_1)\delta I(t_2)\delta I(t_3) \delta
I(t_4)\right\rangle\nonumber\\ && -\tilde{\cal S}_{2}(t_1,t_2)\tilde{\cal
S}_{2}(t_3,t_4)-\tilde{\cal S}_{2}(t_1,t_3)\tilde{\cal
S}_{2}(t_2,t_4)\nonumber\\ && - \tilde{\cal S}_{2}(t_1,t_4)\tilde{\cal
S}_{2}(t_2,t_3).
\end{eqnarray}

In what follows we will use Eq. (\ref{Sn}) as a definition of
the current cumulants also in the quantum case. This definition
unambiguously fixes time ordering of the current operators. The
corresponding expression for the third cumulant ($m=3$) in terms
of the time-ordered current operators has been specified in Ref.
\onlinecite{GGZ}. Analogous expressions for higher cumulants $m>3$
are cumbersome and we do not present them here.

We also define the Fourier transform  of the current cumulants:
\begin{equation}
\tilde {\cal S}_m=\int dt_1\dots dt_{m} \frac{e^{i\omega_1t_1+\dots
+i\omega_{m}t_{m}}}{2\pi}\,\tilde{\cal S}_{m}(t_1,\dots t_m).
\label{Fou}
\end{equation}

We begin with the limit of weakly transmitting barriers
in which case in the lowest order in $T_n$ one can evaluate
the current cumulants
making use of the AES effective action \cite{ESA,SZ}.
Combining Eqs. (\ref{aes2}) and (\ref{Sn}) and dropping the
terms $\propto T_n^2$ for the odd ($m=2l+1$) cumulants one finds
\begin{eqnarray}
&& \tilde{\cal S}_m=\frac{e^{m+1}gV}{2\pi}\,\delta(\omega_1+
\dots+\omega_m),\label{solt}
\end{eqnarray}
where $g=2\sum_n T_n$ is the dimensionless conductance of the scatterer.
Analogously one can evaluate the even ($m=2l$) current cumulants
which read
\begin{eqnarray}
&& \tilde{\cal S}_m=\frac{e^{m}g}{2^{m} \pi}\,
\delta(\omega_1+\dots+\omega_m)\times  \nonumber\\
&&{\sum_{\nu_j=\pm 1}}'\bigg(
eV+\sum_{j=1}^m\frac{\nu_j\omega_j}{2}\bigg)
\coth\bigg(\frac{eV}{2T}+\sum_{j=1}^m\frac{\nu_j\omega_j}{4T}\bigg).\hspace{0.5cm}
\label{selt}
\end{eqnarray}
Here the prime in the sum implies the summation over
``charge" configurations $\nu_j=\pm 1$ with the odd number of positive
(negative) ``charges"  $\nu_j$. This result is also valid in the lowest
order in $T_n$.

Keeping the terms $\propto T_n^2$ in Eq. (\ref{aes2}) and repeating the
same calculation one can evaluate the second order corrections to Eqs.
(\ref{solt}) and (\ref{selt}). The corresponding expressions turn out to
be rather complicated and for this reason are omitted here. We note,
however, that in the limit of low voltages (or high frequencies) $eV\ll
T,\omega_1,\dots ,\omega_m$ the non-local in time terms in the action
(\ref{aes2}) do not contribute to the odd current cumulants. Hence, in
this limit the whole analysis gets much simpler and the odd cumulants are
fully determined by the remaining (local in time) part of the action. In
other words, in the limit $eV\ll T,\omega_1,\dots ,\omega_m$ the odd current
cumulants can be evaluated (up to the terms $\sim T_n^2$) with the aid of
the generating function
\begin{eqnarray}
&& iS_{D}=-\frac{i eV}{\pi}\int_0^{t} dx\;
\sin\varphi^-(x)\times
\nonumber\\ && \left[\sum_nT_n+ \frac{2}{3}\sum_nT_n^2\, \sin^2\frac{\varphi^-
(x)}{2}\right] ,
\end{eqnarray}
which should be substituted into Eq. (\ref{Sn}) instead of $S_{\rm sc}$.

\begin{figure}
\begin{center}
\includegraphics[width=8cm]{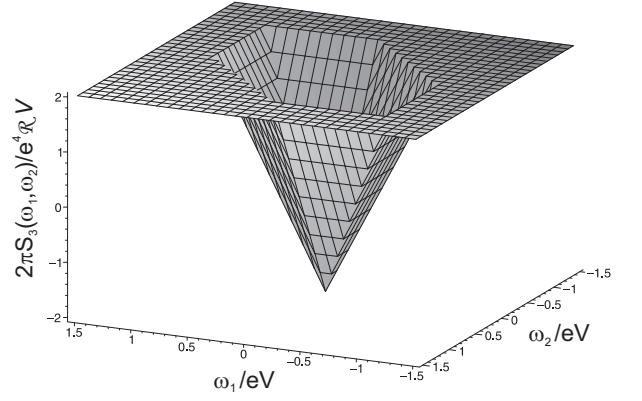}
\end{center}
\caption{Third cumulant of the current operator  ${\cal S}_3$
at zero temperature as a function of frequencies $\omega_1$ and $\omega_2$.
Note that ${\cal S}_3(\omega_1,\omega_2)$ changes its sign with
increasing frequencies and flattens off at $|\omega_1|,|\omega_2|> eV$.}
\end{figure}

Now  we  turn to the case of weakly reflecting barriers, in which case
the effective action is defined by Eqs. (\ref{smallR},\ref{action}).
Combining this action with  Eq. (\ref{Sn}) for the odd ($m=2l+1\geq 3$)
current cumulants we obtain
\begin{eqnarray}
&&\tilde{\cal S}_m=\frac{e^{m}{\cal
R}}{2\pi}\delta(\omega_1+\dots+\omega_m) \bigg[-2eV+ \sum_{\mu_{1,2}=\pm
1}\sum_{\nu_j=\pm 1} \nonumber\\ &&
(-1)^{\frac{\mu_1+\mu_2}{2}}\bigg(eV+\sum_{j=1}^m\frac{\nu_j\omega_j}{2}\bigg)
\coth\bigg(\frac{eV}{2T}+\sum_{j=1}^m\frac{\nu_j\omega_j}{4T}\bigg)
\nonumber\\ &&\times\, \prod_{j=1}^m\bigg(\frac{\mu_1+\mu_2}{4}+\nu_j
\bigg(\frac{\mu_1-\mu_2}{4}-\coth\frac{\omega_j}{2T}\bigg)\bigg) \bigg].
\label{Snodd}
\end{eqnarray}
Analogously for the even ($m=2l \geq 4$) we find
\begin{eqnarray}
&&\tilde{\cal S}_m=-\frac{e^{m}{\cal
R}}{2\pi}\delta(\omega_1+\dots+\omega_m)
\bigg[2\sum_{j=1}^m\omega_j\coth\frac{\omega_j}{2T} \nonumber\\ &&
+\,\sum_{\mu_{1,2}=\pm 1}\sum_{\nu_j=\pm 1}(-1)^{\frac{\mu_1+\mu_2}{2}}
\label{Sneven}\\ &&\times\,
\bigg(eV+\sum_{j=1}^m\frac{\nu_j\omega_j}{2}\bigg)
\coth\bigg(\frac{eV}{2T}+\sum_{j=1}^m
\frac{\nu_j\omega_j}{4T}\bigg)
\nonumber\\ && \times\, \prod_{j=1}^m\bigg(\frac{\mu_1+\mu_2}{4}+\nu_j
\bigg(\frac{\mu_1-\mu_2}{4}-\coth\frac{\omega_j}{2T}\bigg)\bigg) \bigg].
\nonumber
\end{eqnarray}

The delta-functions $\delta(\omega_1+\dots+\omega_m)$ in Eqs.
(\ref{solt},\ref{selt},\ref{Snodd},\ref{Sneven})
illustrate the cumulant invariance with respect
to the time shifts $t_j\to t_j+\Delta t$.
One can also consider ``on-shell'' cumulants ${\cal S}_m$ symbolically
defined as
\begin{eqnarray}
&&{\cal S}_{m}(\omega_1,\dots,\omega_{m-1})= \nonumber\\ &&
=\frac{\tilde{\cal S}_m(\omega_1+ \dots + \omega_{m-1},-\omega_1,-
\omega_2,\dots, -\omega_{m-1})}{\delta(\omega_1+\dots +\omega_m)}.
\label{onshell}
\end{eqnarray}
This definition implies that one should first remove the
$\delta-$function from Eqs.
(\ref{solt},\ref{selt},\ref{Snodd},\ref{Sneven})   and then
replace the frequencies as follows $\omega_1\to \sum_{j=1}^{m-1}\omega_j,$
$\omega_2\to -\omega_1,$ $\omega_3\to -\omega_2,$ ... $\omega_m\to -\omega_{m-1}.$ 
The cumulants ${\cal S}_m$, defined in this way, depend on $m-1$
frequencies and in the limit $\omega_i \to 0$ reduce to the
standard FCS expressions for the current cumulants.

The complete analytical expression
for the third cumulant ${\cal S}_3(\omega_1,\omega_2)$
at arbitrary reflection probabilities
has been previously derived in Ref. \onlinecite{GGZ}
by means of a somewhat different approach. In the limit
of small ${\cal R}$ the result \cite{GGZ} reduces to that
derived here.

The third current cumulant defined by Eqs. (\ref{Snodd},\ref{onshell})
with $m=3$ is plotted in Fig. 1 in the limit $T=0$.
We note that the overall shape of the function ${\cal S}_3(\omega_1,\omega_2)$
remains the same at any transmission values $T_n$
(cf., e.g., Fig. 2 of Ref. \onlinecite{GGZ04})
except for the tunnel junction limit (all $T_n \ll 1$)  in which case no
frequency dispersion of ${\cal S}$ is observed. In a general case
the cumulant ${\cal S}_3$ becomes dispersionless only at
sufficiently high frequencies (or at low voltages) reaching a universal value
\cite{GGZ,GGZ04}  ${\cal S}_3 \to \beta e^4gV/2\pi$.

Higher cumulants depend on larger number of frequencies and their behavior
turns out to be more complicated. As an example in Fig. 2 we plot the fifth
current cumulant (defined by Eqs. (\ref{Snodd},\ref{onshell}) with $m=5$)
at $T=0$ as a function of two frequencies keeping two others fixed.
Unlike the third cumulant, ${\cal S}_5$ does not tend to any universal
frequency-independent value in the high frequency limit.

\begin{figure}
\begin{center}
\includegraphics[width=8cm]{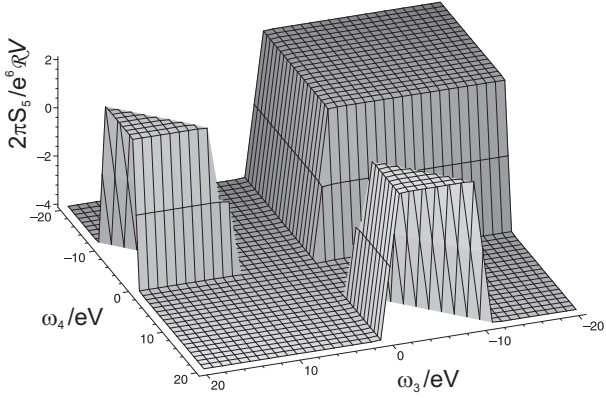}
\end{center}
\caption{Fifth current cumulant ${\cal S}_5$ at $T=0$. Two frequencies
are fixed, $\omega_1=\omega_2=-5eV$, while two others, $\omega_3$ and
$\omega_4$ are varied.}
\end{figure}

\section{Logarithmic corrections and RG}

Let us turn on electron-electron interactions. In this section we assume
the external circuit to be purely Ohmic, $Z_S(\omega)=R_S$. Further we adopt the standard
set of approximations and assume that either the barrier dimensionless
conductance $g=2\sum_n T_n,$ or that of the external circuit $g_S=2\pi
/e^2R_S$ is large, $g\gg 1,$ or $g_S\gg 1$. In this case in a wide range
of parameters the effect of electron-electron interactions is to produce a
negative correction to the system conductance. The latter interaction
correction is proportional to $1/(g+g_S)$ and depends logarithmically
\cite{GZ00,Levy} on voltage and temperature, thus becoming large at
sufficiently small $eV$ and $T$. This divergence demonstrates
insufficiency of the first order perturbation theory in $1/(g+g_S)$ at low
energies and makes it necessary to evaluate the higher order terms.

For the sake of definiteness below in this section
we will assume $T\ll eV$ and define
$$
L=\ln\frac{1}{eVR_0C},
$$
where $R_0=RR_S/(R+R_S)$ and $R=\pi/e^2\sum_nT_n$
is the Landauer resistance of the barrier.
The effective expansion parameter of the perturbation theory
is then $L/(g+g_S).$

Let us first evaluate the second order contribution to the interaction
correction $\propto L^2/(g+g_S)^2$. For this purpose it is sufficient to
expand the effective action to the third order in $\varphi^-.$
One finds \cite{GGZ02}
\begin{eqnarray}
&& iS=iS_S
-\frac{i}{e^2} \int_0^t d
t'\,\varphi^-\left[C\ddot{\varphi}^+ + \frac{1}{R}\dot{\varphi}^+
\right]
\nonumber\\ &&
-\frac{1}{2e^2 R}\int_0^t dt_1dt_2\alpha(t_1-t_2)\varphi^-(t_1)\varphi^-(t_2)
\nonumber\\ && \times\,
\left\{1-\beta +\beta\cos\left[ \varphi^+(t_1)-\varphi^+(t_2)\right]\right\}
\nonumber\\ &&
+\frac{i\beta}{6 e^2 R}\int_0^{t}
d\tau(\varphi^-(\tau))^3\dot\varphi^+(\tau)
\nonumber\\&&
-\frac{2\pi i\gamma}{3e^2 R}\int_0^{t} d\tau_1 d\tau_2 d\tau_3
\varphi^-(\tau_1) \varphi^-(\tau_2)\varphi^-(\tau_3)
\nonumber\\&& \times
f(\tau_2,\tau_1)f(\tau_3,\tau_2)f(\tau_1,\tau_3),
\label{apa}
\end{eqnarray}
where
\begin{equation}
\beta=\frac{\sum_n T_n(1-T_n)}{\sum_n T_n},\;\;
 \gamma=\frac{\sum_n T_n^2(1-T_n)}{\sum_n T_n}
\label{bg}
\end{equation}
and
\begin{equation} f(\tau_2,\tau_1)=\frac{T\sin\left[(\varphi^+(\tau_2)
-\varphi^+(\tau_1))/2 \right]}{\sinh\left[ \pi T(\tau_2-\tau_1)\right]}.
\end{equation}

The current is expressed via the path integral
\begin{eqnarray}
I(t)=\frac{-e\int{\cal D}\varphi^\pm\,\frac{\delta S_{\rm
sc}[\varphi^\pm]}{\delta\varphi^-(t)}\, {\rm e}^{iS[\varphi^\pm]}}
{\int{\cal D}\varphi^\pm\,{\rm e}^{iS[\varphi^\pm]}}.
\label{av}
\end{eqnarray}
This formula can be derived, e.g., from Eq. (\ref{action0})
making use of the definition for the current operator
\begin{equation}
\hat I(t)=\frac{e}{2} \frac{d(\hat N_L(t)-\hat N_R(t))}{dt},
\label{curop}
\end{equation}
where
$\hat N_{L,R}(t)=\sum_{\sigma}\int_{L,R} d^3\bm{r}\,
\hat\Psi^\dagger_\sigma(t,\bm{r})\hat\Psi_\sigma(t,\bm{r})$
is the total number of electrons on the left (right) side of the barrier.

Evaluating this integral with the approximate action (\ref{apa})
and assuming $eVR_0C\lesssim 1$  we get
\begin{eqnarray}
I&=&\frac{V}{R}\left[ 1-\frac{2\beta L}{g+g_S}
+\right.
 \frac{2 L^2}{(g+g_S)^2}\left( \beta -\frac{\beta^2
R_S}{R+R_S}-2\gamma \right)
\nonumber\\ &&
\left.
+\,{\cal O}\left(\frac{L^3}{(g+g_S)^3}\right)\right].
\label{Ilog}
\end{eqnarray}

In order to find the higher order terms of the perturbation theory
in $L/(g+g_S)$ one needs to retain the contributions to the effective
action of order $\sim (\varphi^-)^4$ and higher. An alternative way
is to treat the effective action by means of the RG approach \cite{kn,bn}
which allows to recover the leading logarithmic contributions to all orders.

Previously, the RG equations for this problem were formulated
either in the limit \cite{kn} $g_S\gg 1,g$
or in the opposite limit \cite{bn} $g_S\to 0$ and $g \gg 1$.
In a general case the RG equations for the
channel transmissions read:
\begin{equation}
\frac{d\tilde T_n }{ d L}=-\frac{\tilde T_n(1-\tilde T_n)}{\sum_k \tilde
T_k + (g_S/2)}.
\label{RG}
\end{equation}
The implicit solution for these equations can be obtained in the form
\begin{eqnarray}
 \tilde T_n(L)&=&\frac{T_n (1-z)}{1-T_n z}
\nonumber\\
L&=&-\sum_k\ln(1-T_k z)-\frac{g_S}{2}\ln(1-z),
\label{ren}
\end{eqnarray}
where the parameter $z$ changes from 0 to 1.

Eqs. (\ref{ren}) demonstrate that as the voltage decreases the channels
are ``turned off'' by interactions one by one depending on their
transmission values. Most transparent channels $R_n \ll 1$ remain open
down to lowest voltages. Resolving Eqs. (\ref{ren}) one can explicitly
determine the renormalized (voltage dependent) transmissions $\tilde T_n$
and, substituting $\tilde T_n$ into the Landauer formula, derive the
expression for the $I-V$ curve. Unfortunately it remains unclear whether
this approach is sufficient down to the lowest energies/voltages in which
case instanton effects \cite{Naz,GZepl} need to be taken into account. The
analysis of this problem is beyond the scope of the present paper.

\section{Current in the limit ${\cal R}\ll 1$}

Let us now evaluate the interaction correction to the current for
a somewhat different physical limit. In contrast to the previous
section, here we will make no assumptions about both dimensionless
conductances $g$ and $g_S$, i.e. the conductances of both the
barrier and the external circuit can no longer be large. At the
same time we focus our attention on almost transparent scatterers
assuming ${\cal R}\ll 1$.

We again make use of Eq. (\ref{av}) combining it with the effective action
(\ref{smallR}). In the path integral we perform  a shift
$\varphi^+\to\varphi^++eVt,$ where $V=V_x/(1 + e^2 N_{\rm ch}R_S/\pi )$ is
the voltage drop at the barrier in the absence of interactions. This shift
helps to eliminate the linear in $V_x$ term from the action (\ref{SS}),
but simultaneously introduces the voltage $V$ in the last term of Eq.
(\ref{action}). The variational derivative of the action takes the form:
\begin{eqnarray}
-e\frac{\delta S_{\rm sc}[\varphi^\pm]}{\delta\varphi^-(t)}=
\frac{e\dot\varphi^+(t)}{\pi}(N_{\rm ch}-{\cal R}\cos\varphi^-(t))
\nonumber\\ -\frac{2ie{\cal R}}{\pi}\int dx
\alpha(t-x)\sin\frac{\varphi^-(x)}{2}\cos\frac{\varphi^-(t)}{2} \nonumber\\
\times \cos\bigg[eV(t-x)+\varphi^+(t)-\varphi^+(y) \nonumber\\ +i\int
dz\frac{T\sinh\pi T(t-y)\varphi^-(z)}{\sinh\pi T(t-z)\sinh\pi
T(y-z)}\bigg]+\dots
\label{varI}
\end{eqnarray}
Here $\dots$ stands for the terms which give no contribution to
the current.

Evaluating the path integral (\ref{av}) we can 
put ${\cal R}=0$ in the action $S[\varphi^\pm]$ in Eq.
(\ref{av}) and keep the terms $\propto{\cal R}$ only in $\delta
S_{\rm sc}/\delta\varphi^-$. It is possible because the terms coming from first order in ${\cal R}$ 
correction to the action vanish.
The integral over $\varphi^+$ gives
$\delta-$function which fixes $\varphi^-$. Since the latter phase turns
out to be small, $|\varphi^-|<\pi$, we are allowed to use the action
in the form (\ref{Ssc}). The path integral becomes
Gaussian and can be evaluated exactly. The result reads
\begin{eqnarray}
&& I=\frac{e^2}{\pi}\big(N_{\rm ch}-{\cal R}\big)V \nonumber\\ &&
+\,\frac{2e{\cal R}}{\pi}\int_0^\infty dt\,\alpha(t)\,{
e}^{F(t)}\sin\left[\frac{K(t)}{2}\right]\sin [eVt]. \label{current}
\end{eqnarray}
Here we have defined the functions
\begin{eqnarray}
K(t)=\int\frac{d\omega}{2\pi}\frac{ie^2\,{\rm e}^{-i\omega
t}}{(\omega+i0) \left(-i\omega C+\frac{e^2N_{\rm
ch}}{\pi}+\frac{1}{Z_S(\omega)}\right)}, \label{K}
\\
F(t)=\int\frac{d\omega}{2\pi}\frac{e^2\omega\coth\frac{\omega}{2T}}
{-i\omega C+\frac{e^2N_{\rm ch}}{\pi}+\frac{1}{Z_S(\omega)}}
\frac{1-\cos\omega t}{\omega^2}. \label{F}
\end{eqnarray}
Let us point out that the form of Eq. (\ref{current}) resembles to a certain
extent that of the result for the $I-V$ curve derived perturbatively
in $T_n\ll 1$
for externally shunted tunnel barriers within the so-called $P(E)$-theory
\cite{IN}. Here, in contrast, the interaction term in
Eq. (\ref{current}) was derived perturbatively in
 ${\cal R}=\sum_n(1-T_n)$. Another important feature
of our result is the presence of the term $e^{+F(t)}$ under the time
integral in Eq. (\ref{current}). This exponent becomes large
in the long time limit and should be contrasted with
the decaying exponent $e^{-F(t)}$ in the corresponding expression
\cite{IN} derived in the limit $T_n\ll 1$.

In the limit $V\to\infty$
the $I-V$ dependence (\ref{current}) tends to the following simple form
\begin{equation}
I=\frac{e^2}{\pi}\big(N_{\rm ch}-{\cal R}\big)V-\frac{e^2}{\pi}{\cal R}\frac{e}{2C},
\end{equation}
i.e.  the $I-V$ curve has the offset. This result formally holds for
any $Z_S(\omega)$ and at any temperature. In practice, the offset might be
difficult to observe at high conductances and temperature.

Below we will consider an important limit of purely Ohmic
external environment $Z_S(\omega)=R_S$. If both temperature and
voltage are sufficiently small, $T,eV\ll (R+R_S)/RR_SC$, the
integral in Eq. (\ref{current}) can be evaluated analytically. We
obtain
\begin{eqnarray}
I&=&\frac{e^2N_{\rm ch}}{\pi}V -\left(\frac{{\rm
e}^{\gamma_0}(R+R_S)}{2\pi TRR_SC}\right)^{\frac{2}{g+g_S}}
\nonumber\\ &&\times\, \frac{4\pi eT{\cal
R}}{\Gamma\left(2-\frac{2}{g+g_S}\right)\left|\Gamma\left(\frac{1}{g+g_S}+i\frac{eV}{2\pi
T}\right)\right|^2} \nonumber\\ &&\times\,
\frac{\sinh\frac{eV}{2T}}{\cosh\frac{eV}{T}-\cos\frac{2\pi}{g+g_S}},
\label{current1}
\end{eqnarray}
where $\Gamma (x$) is the gamma-function and $\gamma_0 \simeq 0.577$ is the
 Euler constant. At low temperatures, $T\ll eV$, Eq. (\ref{current1})
yields the differential conductance
\begin{equation}
\frac{dI}{dV}=\frac{e^2N_{\rm ch}}{\pi}-\frac{e^2{\cal
R}}{\pi\Gamma\left(1-\frac{2}{g+g_s}\right)} \left(\frac{{\rm
e}^{\gamma_0}(R+R_S)}{e|V|RR_SC}\right)^{\frac{2}{g+g_s}},
\label{didv}
\end{equation}
while in the zero bias limit $eV \ll T$ one recovers the linear
conductance
\begin{equation}
G=\frac{e^2N_{\rm ch}}{\pi}-\frac{e^2{\cal R}}{2\sqrt{\pi}}
\frac{\Gamma\left(1-\frac{1}{g+g_s}\right)}{\Gamma\left(\frac{3}{2}-\frac{1}{g+g_s}\right)}
\left(\frac{{\rm e}^{\gamma_0}(R+R_S)}{\pi
TRR_SC}\right)^{\frac{2}{g+g_s}}. \label{G}
\end{equation}
We note that Eqs. (\ref{current}-\ref{G}), being perturbative in
${\cal R}$,  become inapplicable at voltages/temperatures below
the energy scale $E^*$, which can be estimated as
\begin{equation}
E^*=\frac{R+R_S}{RR_SC}\,\left(\frac{{\cal R}}{N_{\rm
ch}}\right)^{\frac{g+g_s}{2}}. \label{E*}
\end{equation}
At the same time the above results are non-perturbative in both
$1/g$ and $1/g_S$ and, hence, provide complementary information to
the results obtained from the RG analysis discussed in the previous section.

\section{Current noise}

We now turn to the current noise and again consider the system
with almost perfectly transmitting channels assuming ${\cal R} \ll 1$.
The current noise spectrum is defined in a standard manner:
\begin{equation}
{\cal S}_I(\omega)=\int dt\,{\cal S}(t)\,{\rm e}^{i\omega t},
\end{equation}
where
\begin{equation}
{\cal S}(t)=\langle \hat I(t)\hat I(0)+\hat I(0)\hat I(t)\rangle
-2\langle I(0)\rangle^2.
\end{equation}

In the absence of interactions
and in the limit ${\cal R} \ll 1$ the low frequency
noise spectrum scales as \cite{Khlus} $ {\cal S}_I(0 ) \propto {\cal
  R}$. Below we will demonstrate that electron-electron interactions
yield an additional contribution to  $ {\cal S}_I(0 )$ which is also
proportional to  $ {\cal R}$ but can be much larger than the non-interacting
result. In other words, a dramatic {\it increase} of the current shot noise
by electron-electron interactions is expected provided
both conductances $g$ and $g_S$ are not very large.

The noise spectrum will be evaluated with the aid of the path integral
\begin{eqnarray}
{\cal S}_I(\omega)=-2ie^2\int dt\,{\rm e}^{i\omega t}
\frac{\int{\cal D}\varphi^\pm\,\frac{\delta^2 S_{\rm
sc}[\varphi^\pm]}{\delta\varphi^-(t)\delta\varphi^-(0)} \;{\rm
e}^{iS[\varphi^\pm]}} {\int{\cal D}\varphi^\pm\,{\rm
e}^{iS[\varphi^\pm]}}. \label{SIdef}
\end{eqnarray}
This expression follows directly from Eqs. (\ref{action0},\ref{curop}).

The variational derivative of the effective action  $S_{\rm sc}$ can be
evaluated in a straightforward manner. We obtain
\begin{widetext}
\begin{eqnarray}
 && -2ie^2\frac{\delta^2 S_{\rm
sc}[\varphi^\pm]}{\delta\varphi^-(t)\delta\varphi^-(0)}= \frac{2ie^2{\cal
R}}{\pi}\,\dot\varphi^+(t)\sin[\varphi^-(0)]\delta(t)+\frac{2e^2N_{\rm
ch}}{\pi}\alpha(t)- \frac{2e^2{\cal
R}}{\pi}\alpha(t)\big(\cos[\varphi^-(t)]+\cos[\varphi^-(0)]\big)
\nonumber\\ &&
+\frac{2e^2{\cal R}}{\pi}\,\alpha(t)\cos\left[\varphi^+(t)-\varphi^+(0)-i\int
dz\big(T\coth[\pi T(t-z)]+T\coth[\pi Tz]\big)\varphi^-(z)\right]
\cos\frac{\varphi^-(t)}{2}\cos\frac{\varphi^-(0)}{2}
\nonumber\\ &&
+\frac{4ie^2{\cal R}}{\pi}\int dx\,\alpha(t-x)
\sin\left[\varphi^+(t)-\varphi^+(x)-i\int dz\big(T\coth[\pi
T(t-z)]+T\coth[\pi T(z-x)]\big)\varphi^-(z)\right]
\nonumber\\ &&
\times
\big(T\coth[\pi Tt]-T\coth[\pi
Tx]\big)\cos\frac{\varphi^-(t)}{2}\sin\frac{\varphi^-(x)}{2}
\nonumber\\ &&
+\frac{4ie^2{\cal R}}{\pi}\int dx\,\alpha(-x)
\sin\left[\varphi^+(x)-\varphi^+(0)-i\int dz\big(T\coth[\pi
T(x-z)]+T\coth[\pi Tz]\big)\varphi^-(z)\right]
\nonumber\\ &&
\times
\big(T\coth[\pi T(x-t)]-T\coth[\pi
Tt]\big)\cos\frac{\varphi^-(0)}{2}\sin\frac{\varphi^-(x)}{2} +\dots.
\label{d2S}
\end{eqnarray}
\end{widetext}
Here we again omit terms which contribution to the path integral vanishes.

Since we are going to evaluate linear in ${\cal R}$ contributions to
the noise spectrum, it suffices to set ${\cal R}=0$ in the
expression for $S$ in the exponent of Eq. (\ref{SIdef}). As in the previous
section integrating first over $\varphi^+$
one can verify that only the values $|\varphi^-|<\pi$ give a non-vanishing
contribution. Under this
approximation the path integral becomes Gaussian and we find
\begin{eqnarray}
&& {\cal S}_I(t)=\frac{e^4{\cal R}}{\pi C}\delta(t)
+\frac{2e^2(N_{\rm ch}-2{\cal R})}{\pi}\alpha(t) \nonumber\\ &&
+\,\frac{2e^2{\cal R}}{\pi}\alpha(t)\,{\rm
e}^{F(t)}\cos\left[\frac{K(t)}{2}\right]\cos[eVt] \nonumber\\ &&
+\,\frac{4e^2{\cal R}}{\pi}\int dx\,\alpha(x)\,{\rm
e}^{F(x)}\sin\left[\frac{K(x)}{2}\right]\cos[eVx] \nonumber\\
&&\times\, \big(T\coth[\pi T(x-t)]+T\coth[\pi T(x+t)]\big).
\label{SI1}
\end{eqnarray}

Let us define the function
\begin{eqnarray}
\Phi(t)&=&F(t)+\frac{i}{2}K(|t|){\rm sign} t \nonumber\\
&=&e^2\int\frac{d\omega}{2\pi}\frac{(1-\cos\omega
t)\coth\frac{\omega}{2T}+i\sin\omega t} {\omega\left(-i\omega
C+\frac{e^2N_{\rm ch}}{\pi}+\frac{1}{Z_S(\omega)}\right)}, \label{Phi}
\end{eqnarray}
and express the noise power spectrum as follows:
\begin{eqnarray}
&& {\cal S}_I(\omega)=\frac{e^4{\cal R}}{\pi
C}+\frac{2e^2}{\pi}(N_{\rm ch}-2{\cal
R})\omega\coth\frac{\omega}{2T} \nonumber\\ &&
 +\,\frac{2e^2{\cal
R}}{\pi}\int dx\,\alpha(x)\,{\rm e}^{\Phi(x)}\cos[eVx]\cos\omega x
\nonumber\\ &&
-\,\frac{4ie^2{\cal R}\coth\frac{\omega}{2T}}{\pi}\int dx\,\alpha(x)\,{\rm
e}^{\Phi(x)}\cos[eVx]\sin\omega x.
\hspace{0.5cm}
\label{SI2}
\end{eqnarray}
Making use of the property $\Phi(t-i/T)=\Phi(-t)$
one can prove the following identity:
\begin{eqnarray}
\int dx\,\alpha(x)\,{\rm e}^{\Phi(x)}\cos\omega
x=-\frac{e^2}{2C}+\omega\coth\frac{\omega}{2T} \nonumber\\
+\,i\coth\frac{\omega}{2T}\int dx\,\alpha(x)\,{\rm e}^{\Phi(x)}\sin\omega x.
\label{id}
\end{eqnarray}
Then, combining Eqs. (\ref{current}), (\ref{SI2}) and (\ref{id}) we obtain
\begin{eqnarray}
 {\cal S}_I(\omega)&=&\frac{2e^2(N_{\rm ch}-2{\cal
R})}{\pi}\,\omega\coth\frac{\omega}{2T}+ \nonumber\\ &&
+\, \frac{e^2{\cal R}}{\pi}\sum_\pm(\omega\pm eV)\coth\frac{\omega\pm eV}{2T}
\nonumber\\ &&
+\,e\sum_\pm \bigg[2\,\delta I\left(\frac{\omega}{e}\pm V\right)\coth\frac{\omega}{2T}
\nonumber\\ &&
-\,\delta
I\left(\frac{\omega}{e}\pm V\right)\coth\frac{\omega\pm eV}{2T}\bigg],
\label{noise}
\end{eqnarray}
where $\delta I(V)=I(V)-\frac{e^2(N_{\rm ch}-{\cal R})}{\pi}V$ is
the interaction correction to the current, $I(V)$ is defined in
Eq. (\ref{current}). The frequency dependence of the noise
spectrum is illustrated in Fig. 3. One can see that the noise is
enhanced due to  interaction at low frequencies $|\omega|\lesssim
eV,$ and reduced at $|\omega|\gtrsim eV.$

\begin{figure}
\begin{center}
\includegraphics[width=8cm]{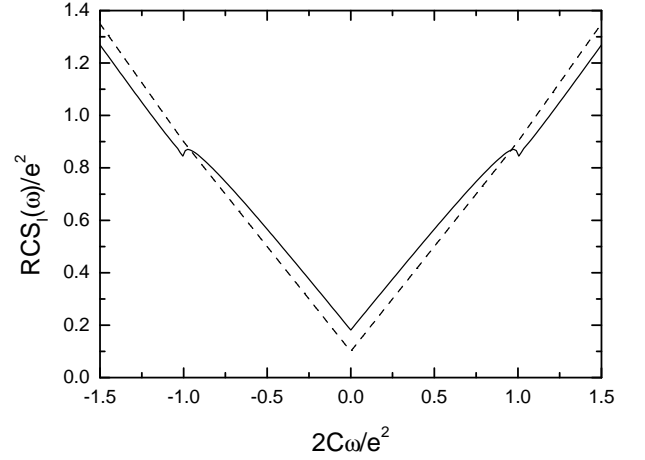}
\end{center}
\caption{Frequency dependence of the noise power spectrum (solid line)
at $T=0$ and in the presence of interactions.
For comparison the noise spectrum in the absence of interactions is shown
by the dashed line. Both curves are plotted for
$N_{\rm ch}=1,$ ${\cal R}=0.1$, $g_S=2$ and $V=e/2C.$
}
\end{figure}

It is easy to see that Eq. (\ref{noise}) satisfies the
fluctuation-dissipation theorem (FDT). Indeed, applying an ac bias,
$V(t)=V_0\cos\omega t,$ and repeating the analysis of the previous section
one can verify that the real part of the  zero-bias conductance is related
to the current in the following way:
\begin{equation}
{\rm Re} \,G(\omega)=\frac{e}{\omega} I\left(\frac{\omega}{e}\right).
\end{equation}
Making use of this identity and setting $V=0$ in Eq.
(\ref{noise}), one finds
\begin{equation}
{\cal S}_I(\omega,V=0)=2\left[{\rm
Re}\,G(\omega)\right]\omega\coth\frac{\omega}{2T}
\end{equation}
in agreement with FDT.

In the non-interacting limit Eq. (\ref{noise}) reduces to the Khlus
formula\cite{Khlus} expanded in ${\cal R}$ to the first order. We also
note that the result for the noise spectrum for highly transmitting
barriers (\ref{noise}) is to some extent similar to that for the case of
tunnel junctions. The latter can be expressed in terms of the $I-V$
dependence in the following way:
\begin{eqnarray}
S_I^{\rm tun}(\omega)=e\sum_\pm I\left(\frac{\omega}{e}\pm eV
\right)\coth\frac{\omega\pm eV}{2T}.
\label{noisetunnel}
\end{eqnarray}
This is a general result which holds in the non-interacting limit \cite{Dahm}
as well as in the presence of an arbitrary external  impedance \cite{Lee}.
We observe that in both Eqs. (\ref{noise}) and (\ref{noisetunnel}) the
effect of electron-electron interactions is fully described
by the interaction term contained in the $I-V$ curve.

\begin{figure}
\begin{center}
\includegraphics[width=8cm]{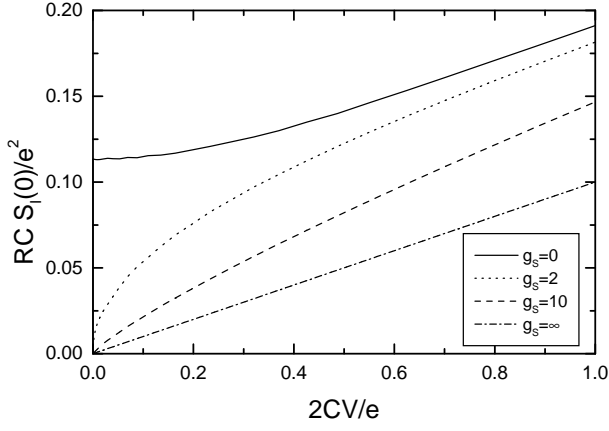}
\end{center}
\caption{The zero frequency shot noise power spectrum ${\cal S}_I(0)$ as a
  function
of voltage $V$ at $T=0$ for $N_{\rm ch}=1$, $R=\pi /e^2$ and ${\cal R}=0.1$.
The conductance $g_S$  effectively controls the interaction
strength, i.e. at small $g_S$ the interaction is strong,
while it tends to zero in the limit $g_S\to\infty$. The interaction-induced
excess noise is clearly observed even at rather large values of
$g_S$.}
\end{figure}

Consider now the low frequency limit $\omega \to 0.$ From the above results
we obtain
\begin{eqnarray}
{\cal S}_I(0)&=&\left(\frac{e^2}{\pi}(N_{\rm ch}-2{\cal
R})+2\frac{d}{dV}\delta I(V)\right)4T \nonumber\\ && +\,
2e\left(\frac{e^2{\cal R}}{\pi}V-\delta
I(V)\right)\coth\frac{eV}{2T}.
\end{eqnarray}
In the shot noise regime $T\ll eV$ this formula reduces to
\begin{eqnarray}
{\cal S}_I(0)=2e\left(\frac{e^2{\cal R}}{\pi}V-\delta I(V)\right).
\label{shn}
\end{eqnarray}
Since the interaction correction is negative, $\delta I(V)<0,$
we arrive at the conclusion (cf. Ref. \onlinecite{GGZ02}) that for highly
transmitting barriers the shot noise is enhanced by electron-electron
interactions. At high voltages the noise spectrum takes the form
\begin{eqnarray}
{\cal S}_I(0)=\frac{2e^3{\cal R}}{\pi}\left(V+\frac{e}{2C}\right),
\label{offset}
\end{eqnarray}
i.e. in this limit interactions induce voltage-independent excess noise
which  magnitude scales linearly with the Coulomb gap $e/2C$.

In the limit $eVRR_SC/(R+R_S)\ll 1$ from Eqs. (\ref{shn}) and
(\ref{current1}) one obtains the result
\begin{eqnarray}
{\cal S}_I(0)=\frac{2e^2{\cal R}(R+R_S){\rm
e}^{\frac{2\gamma_0}{g+g_s}}}{\pi\Gamma\left(2-\frac{2}{g+g_s}\right)RR_SC}
\left(\frac{e|V|RR_SC}{R+R_S}\right)^{1-\frac{2}{g+g_s}}.
\label{shot}
\end{eqnarray}
This expression is valid provided $e|V| > E^*$, where $E^*$ is
defined in Eq. (\ref{E*}). In the limit $g+g_S \to \infty$ Eq.
(\ref{shot}) reduces to the non-interacting result. For $g+g_S\gg
1$ the interaction effects remain weak and Eq. (\ref{shot}) agrees
with the perturbative results \cite{GGZ02}. In the limit $g+g_S\gg
1$ the same expressions can also be reproduced with the aid of the
RG analysis (\ref{RG},\ref{ren}) combined with the formula
$S_I=\frac{2e^2}{\pi} eV\sum_n\tilde T_n(1-\tilde T_n)$. For
smaller conductance values the excess noise becomes larger and for
$g+g_S$ of order one the shot noise is strongly enhanced by
electron-electron interactions. The effect of interactions on the
shot noise spectrum ${\cal S}_I (0)$ of highly transmitting
coherent scatterers is clearly observed in Fig. 4.

\section{Quantum dots}

So far we have considered a single scatterer embedded in
the electromagnetic environment with the impedance $Z_S(\omega)$.
Another important physical situation is that of so-called quantum dots.
A quantum dot can be modeled by a system of two scatterers connected in
series via a sufficiently small island. In this section we will demonstrate
that some of the results derived above for a single scatterer can be directly
generalized to the case of quantum dots.

Let us denote the number of channels in the left and right barriers $N_l$
and $N_r.$ As before, we will also assume that all these channels are
almost open and ${\cal R}_l,{\cal R}_r\ll 1.$ Throughout our analysis we
will stick to a simplified description which amounts to treating the
effect of the left barrier on the right one (and vice versa) as that of an
Ohmic resistor. This approximation is justified if the electron
distribution function inside the dot remains in equilibrium, i.e.
equal to the Fermi function with the temperature of the leads. This is the
case either in the linear in voltage regime or, else, provided inelastic
relaxation of electrons inside the dot is sufficiently strong. The latter
condition may apply for large quantum dots and/or at high enough
temperatures. In addition, below we will ignore the gate
modulation of the current flowing across the quantum dot.

Let us denote the voltage drop across the left/right barrier as $V_{l,r}$.
Obviously, $V_l+V_r=V,$ where $V$ is the total bias voltage.
From the current conservation conditions at both barriers one finds
\begin{eqnarray}
I=\frac{e^2(N_l-{\cal R}_l)}{\pi}V_l+\delta I_l(V_l), \nonumber\\
I=\frac{e^2(N_r-{\cal R}_r)}{\pi}V_r+\delta I_r(V_r).
\label{cons}
\end{eqnarray}
Here $\delta I_{l}$ is the interaction correction to the current
in the left barrier given by the last term in Eq. (\ref{current})
and Eqs. (\ref{K},\ref{F}) with the following replacements:
$N_{\rm ch}\to N_l$,
$1/Z_S(\omega)\to (\frac{e^2}{\pi})N_r$, $C\to C_\Sigma$, where $C_\Sigma$
is the total capacitance of the quantum dot. The interaction correction
 $\delta I_{r}$ is defined analogously.

Resolving Eqs. (\ref{cons}) and keeping  the first order in
${\cal R}_{l,r}$, one finds
\begin{eqnarray}
I&=&\frac{e^2}{\pi}\frac{N_lN_r}{N}V
-\frac{e^2}{\pi}\frac{N_r^2{\cal R}_l+N_l^2{\cal R}_r}{N^2}V
\nonumber\\ &&
+\,\frac{N_r}{N}\delta I_l\left(\frac{N_r}{N}V\right)
+\frac{N_l}{N}\delta I_r\left(\frac{N_l}{N}V\right),
\end{eqnarray}
where $N=N_l+N_r$. Assuming now
$T\ll eV$ and $eVR_lR_rC_\Sigma/(R_l+R_r)\ll 1$ we
obtain
\begin{eqnarray}
&& I=\frac{e^2}{\pi}\frac{N_lN_r}{N}V \nonumber\\ && -\,\frac{2{\rm
e}^{\gamma_0/N}eE_C(N_r^{2-{1}/{N}}{\cal R}_l+N_l^{2-1/N}{\cal
R}_r)}{\pi^2N^{2-2/N}\Gamma\left(2-\frac{1}{N}\right)} \left(\frac{\pi
eV}{2E_C}\right)^{1-1/N}
\end{eqnarray}
and
\begin{eqnarray}
&& \frac{dI}{dV}=\frac{e^2}{\pi}\frac{N_lN_r}{N} \nonumber\\ &&
-\,\frac{e^2}{\pi}\frac{N_r^{2-1/N}{\cal R}_l+N_l^{2-1/N}{\cal
R}_r}{N^{2-2/N}\Gamma\left(2-\frac{1}{N}\right)} \left(\frac{2\,{\rm
e}^{\gamma_0} E_C}{\pi eV}\right)^{1/N},
\end{eqnarray}
where $E_C=e^2/2C_\Sigma$ is the charging energy.
In the opposite limit $eV \ll T\ll (R_l+R_r)/R_lR_rC_\Sigma$ we arrive at the expression
for the  linear conductance
\begin{eqnarray}
&& G=\frac{e^2}{\pi}\frac{N_lN_r}{N} \nonumber\\ &&
-\,\frac{e^2}{2\sqrt{\pi}}\frac{\Gamma\left(1-\frac{1}{2N}\right)(N_r^2{\cal
R}_l+N_l^2{\cal R}_r)} {\Gamma\left(\frac{3}{2}-\frac{1}{2N}\right)
N^{2-1/N}} \left(\frac{2\,{\rm e}^{\gamma_0}E_C}{\pi^2 T}\right)^{1/N}.
\label{conddot}
\end{eqnarray}
In the limit $N_l=N_r=1$ this result reduces to that derived
in Ref. \onlinecite{Matveev} by means of the bosonization technique.
Eq. (\ref{conddot}) is also consistent with the corresponding result of
Ref. \onlinecite{ABG} in the case of arbitrary number of channels $N_{l,r}$.

Under the assumption of strong inelastic relaxation inside the dot
current fluctuations across two barriers can be considered uncorrelated and
one can also evaluate the noise spectrum of the system. Combining Eqs. $\delta
I=\delta V_l/R_l+\xi_l,$ $\delta I=\delta V_r/R_r+\xi_r$ and $\delta
V_l+\delta V_r=0$ we arrive at the relation between the current fluctuations
in the system $\delta I$ and fluctuations across each of the barriers $\xi_{l,r}$:
\begin{equation}
\delta I=\frac{N_r\xi_l+N_l\xi_r}{N_l+N_r}.
\end{equation}
Accordingly, the noise power spectrum
${\cal S}_I\propto \langle\delta I^2\rangle$ takes the form
\begin{equation}
{\cal S}_I=\frac{N_r^2}{N^2}{\cal S}_l+\frac{N_l^2}{N^2}{\cal
S}_r,
\end{equation}
where the noise spectra ${\cal S}_{l,r}$ are defined by Eq.
(\ref{shot}) with the corresponding parameters. Here we consider
only the shot noise regime $T\ll eV$ and assume $eV\ll NE_C,$ in
which case one finds
\begin{eqnarray}
{\cal S}_I=\frac{4\,{\rm e}^{\gamma_0/N}(N_r^2{\cal R}_l+N_l^2{\cal
R}_r)}{\pi^2N^{2-1/N}\Gamma\left(2-\frac{1}{N}\right)}
e^2E_C\left(\frac{\pi eV}{2E_C}\right)^{1-1/N}.
\end{eqnarray}
As before, we observe strong enhancement of the shot noise by
electron-electron interactions provided the number of channels in the
dot $N$ is not large.

Further work is needed in order to relax the assumption about strong
inelastic relaxation inside the dot. In the metallic limit
$g \gg 1$ the corresponding analysis has been developed in Ref.
\onlinecite{GZ03}.

\centerline{\bf Acknowledgments}

We acknowledge stimulating discussions with D.A. Bagrets and S.V. Sharov.
This work is part of the Kompetenznetz ``Funktionelle
Nanostructuren'' supported by the Landestiftung
Baden-W\"urttemberg gGmbH and of the European Community's Framework Programme
NMP4-CT-2003-505457 ULTRA-1D "Experimental and theoretical investigation of
electron transport in ultra-narrow 1-dimensional nanostructures".

\appendix

\section{Transformation of the action}

Let us demonstrate the equivalence of the actions (\ref{action1})
and (\ref{action2}). 
First we evaluate the Green function $\check G_V$ which satisfies
the equation
\begin{equation}
\check G_V^{-1}(t_1,\bm{r}_1)\check
G_V(t_1,t_2;\bm{r}_1,\bm{r}_2)=\delta(t_1-t_2)\delta(\bm{r}_1-\bm{r}_2),
\label{A1}
\end{equation}
where $\check G_V^{-1}(t,\bm{r})$ is defined in Eq. (\ref{GV-1}).
The general solution of Eq. (\ref{A1}) can be written in the form
\begin{eqnarray}
\check G_V=\left(\begin{array}{cc}
-i\theta(t_1-t_2){\cal U}_1(t_1,t_2) & 0 \\
0 & i\theta(t_2-t_1){\cal U}_2(t_1,t_2)
\end{array}\right)
\nonumber\\
+\,i\left(\begin{array}{cc}
{\cal U}_1(t_1,t) & 0 \\
0 & {\cal U}_2(t_1,t)
\end{array}\right)
\left(\begin{array}{cc}
\hat\rho_V(t) & -\hat\rho_V(t) \\
\hat\rho_V(t)-1 & -\hat\rho_V(t)
\end{array}\right)
\nonumber\\ \times\,
\left(\begin{array}{cc}
{\cal U}_1(t,t_2) & 0 \\
0 & {\cal U}_2(t,t_2)
\end{array}\right).\hspace{3.5cm}
\label{GV1}
\end{eqnarray}
Here $\hat\rho_V(t)$ is an arbitrary operator. Below we will use
the convention according to which the product of operators
involves only the coordinate integration, i.e.: $ {\cal
U}^2(t_1,t_2,\bm{r}_1,\bm{r_2})=\int d^3\bm{r}'{\cal
U}(t_1,t_2,\bm{r}_1,\bm{r}') {\cal
U}(t_1,t_2,\bm{r}',\bm{r}_2).$ It is also important to keep in
mind that $0<t_1,t_2<t,$ and the Green function $\check G_V$
implicitly depends on the final time $t$.

The task at hand is to evaluate the operator $\hat\rho_V(t).$ For
this purpose let us make use of the Dyson equation
\begin{eqnarray}
\check G_V(t_1,t_2)=\check G_0(t_1,t_2)\hspace{3.8cm}
\nonumber\\
-\, \int_0^t dt'\check G_0(t_1,t')\, e\check V(t')\,\check G_V(t',t_2),
\label{Dyson}
\end{eqnarray}
where
\begin{equation}
\check V(t)=\left(\begin{array}{cc}
V_1(t) & 0 \\
0 & V_2(t)
\end{array}\right),
\end{equation}
and $\check G_0$ is the Keldysh-Green function of noninteracting
electrons:
\begin{eqnarray}
\check G_0=\left(\begin{array}{cc}
-i\theta(t_1-t_2){\cal U}_0(t_1,t_2) & 0 \\
0 & i\theta(t_2-t_1){\cal U}_0(t_1,t_2)
\end{array}\right)
\nonumber\\
+\,i\left(\begin{array}{cc}
{\cal U}_0(t_1,0) & 0 \\
0 & {\cal U}_0(t_1,0)
\end{array}\right)
\left(\begin{array}{cc}
\hat\rho_0 & -\hat\rho_0 \\
\hat\rho_0-1 & -\hat\rho_0
\end{array}\right)
\nonumber\\ \times\,
\left(\begin{array}{cc}
{\cal U}_0(0,t_2) & 0 \\
0 & {\cal U}_0(0,t_2)
\end{array}\right).\hspace{3.5cm}
\label{G0}
\end{eqnarray}
Here ${\cal U}_0$ is the evolution operator for
non-interacting electrons defined by eq. (\ref{U12}) with
$V_{1,2}=0$ and $\hat \rho_0$ is the initial reduced single-electron
density matrix operator defined as
$\langle \bm{r}|\hat\rho_0|\bm{r}'\rangle=
{\rm
  tr}\left(\hat\Psi^\dagger_\uparrow(\bm{r}')\hat\Psi_\uparrow(\bm{r})
  \bm{\hat\rho}_0 \right)$, where $\bm{\hat\rho}_0$ is the initial
  many-particle density matrix. Substituting the general solution
(\ref{GV1}) into Eq. (\ref{Dyson}) and making use of the Dyson
equation for the evolution operators
$$
{\cal U}_{1,2}(t,t')={\cal U}_{0}(t,t')
+i\int^t_{t'} ds\,{\cal U}_{0}(t,s)eV_{1,2}(s)
{\cal U}_{1,2}(s,t'),
$$
 we find\cite{GZ1}:
\begin{eqnarray}
\hat\rho_V(t)={\cal U}_1(t,0)\hat\rho_0\hspace{4cm}
\nonumber\\ \times\,
\left[1+({\cal U}_2(0,t){\cal U}_1(t,0)-1)
\hat\rho_0\right]^{-1}{\cal U}_2(0,t).
\label{rhoV}
\end{eqnarray}

As a next step, let us fulfill the following replacements $eV_1\to
eV^+ +\lambda V^-/2,$ $eV_2\to eV^+ -\lambda V^-/2 $ and evaluate
the derivative $i\partial S_{el}/\partial\lambda,$ where
$iS_{el}=2{\rm Tr}\ln\check G_V^{-1}.$ With the aid of Eq.
(\ref{GV-1}) we obtain
\begin{eqnarray}
i\frac{\partial S_{el}}{\partial\lambda}=\int_0^t ds\int d^3\bm{r}
\left(G_{V,\lambda}^{11}(s,s,\bm{r},\bm{r})
\right.
\nonumber\\
\left.
-G_{V,\lambda}^{22}(s,s,\bm{r},\bm{r})\right)eV^-(s,\bm{r}).
\end{eqnarray}
Employing Eqs. (\ref{GV1}), (\ref{rhoV}) and using the properties
of the trace of the product of operators, we find
\begin{eqnarray}
&& i\frac{\partial S_{el}}{\partial\lambda} =i\int_0^t ds\,{\rm
tr}\left[\big({\cal U}_{1,\lambda}(s,t) \hat\rho_{V,\lambda}(t){\cal
U}_{1,\lambda}(t,s) \right. \nonumber\\ && \left. +\, {\cal
U}_{2,\lambda}(s,t)\hat\rho_{V,\lambda}(t) {\cal
U}_{2,\lambda}(t,s)\big)\,e\hat V^-(s)\right] \nonumber\\ &=&i\int_0^t
ds\, \nonumber\\ && \times\, {\rm tr}\left[\big({\cal U}_{2,\lambda}(0,t)
{\cal U}_{1,\lambda}(t,s)e\hat V^-(s){\cal U}_{1,\lambda}(s,0)\right.
\nonumber\\ && +\,{\cal U}_{2,\lambda}(0,s)e\hat V^-(s){\cal
U}_{2,\lambda}(s,t) {\cal U}_{1,\lambda}(t,0)  \big) \nonumber\\ &&
\times\,\left. \hat\rho_0 [1+(\hat{\cal U}_{2,\lambda}(0,t)\hat{\cal
U}_{1,\lambda}(t,0)-1)\hat\rho_0]^{-1}\right]. \label{dSdlambda}
\end{eqnarray}

What remains is to integrate Eq. (\ref{dSdlambda}) over $\lambda$
from 0 to 1. This task is accomplished with the aid of the
identity
\begin{eqnarray}
&&\frac{\partial}{\partial\lambda}
\left({\cal U}_{2,\lambda}(0,t){\cal U}_{1,\lambda}(t,0)\right)=
\nonumber\\ &&
\frac{i}{2}\int_0^t ds \big({\cal U}_{2,\lambda}(0,t)
{\cal U}_{1,\lambda}(t,s)e\hat V^-(s){\cal U}_{1,\lambda}(s,0)
\nonumber\\ &&
+\,{\cal U}_{2,\lambda}(0,s)e\hat V^-(s)
{\cal U}_{2,\lambda}(s,t){\cal U}_{1,\lambda}(t,0)  \big).
\end{eqnarray}
Since the action $S$ equals to zero at $\lambda=0,$ we arrive at
the final result
\begin{eqnarray}
2{\rm Tr}\ln\check G_V^{-1}=2{\rm tr}\ln
\left[1+({\cal U}_{2}(0,t){\cal U}_{1}(t,0)-1)\hat\rho_0\right],
\end{eqnarray}
which proves the equivalence of Eqs. (\ref{action1}) and (\ref{action2}).

\section{Fully transmitting barriers}

Let us demonstrate that the action $S_0$ (\ref{S0}) can be exactly
transformed to the form (\ref{s0}). For this purpose
we first expand the action (\ref{S0}) in the following series
\begin{eqnarray}
iS_0[a,\varphi^+]&=& 2{\rm Tr}\left\{ a(x)\rho_0(y-x){\rm
e}^{i\frac{\varphi^+(y)-\varphi^+(x)}{2}}\right\} \nonumber\\ &&
-2\sum_{N=2}^\infty \frac{(-1)^NJ_N}{N} , \label{S01}
\end{eqnarray}
where
\begin{eqnarray}
J_N&=&\int dx_1\dots dx_N\,
\rho_0(x_1-x_2)a(x_2)\rho_0(x_2-x_3)\dots \nonumber\\ &&\times\,
\rho_0(x_{N-1}-x_{N})a(x_N)\rho_0(x_N-x_1)a(x_N), \label{JN}
\end{eqnarray}
and rewrite $J_N$ in the form
\begin{eqnarray}
&& J_N=\int dx\int_{-E_{\rm min}}^\infty\frac{dE}{2\pi}
\int\frac{d\omega_1\dots
d\omega_{N}}{(2\pi)^{N}}\,n(E)
\nonumber\\ && \times\,
n(E+\omega_1)n(E+\omega_1+\omega_2)\dots n(E+\omega_1+\dots+\omega_{N-1})
\nonumber\\ &&
\times\, \tilde a(\omega_1)\tilde a(\omega_2)\dots \tilde
a(\omega_{N})\,{\rm e}^{i(\omega_1+\dots+\omega_N)x}, \hspace{0.5cm}
\label{JN11}
\end{eqnarray}
where $\tilde a(\omega)=\int dx\,{\rm e}^{i\omega x}a(x)$ is the
Fourier transform of $a(x),$ and we have also introduced the
cutoff at a large negative energy $-E_{\rm min}$.

We will use the following property of the Fermi function
\begin{eqnarray}
n(E)n(E+\omega)=\frac{1}{2}\left(1-\coth\frac{\omega}{2T}\right)n(E)
\nonumber\\
+\,\frac{1}{2}\left(1+\coth\frac{\omega}{2T}\right)n(E+\omega).
\label{prop0}
\end{eqnarray}
This equation is analogous to the Ward identity for the electron
Matsubara Green function: ${\cal G}(\omega+\Omega,p+q){\cal
G}(\omega,p)= {\cal G}(\Omega,q)[{\cal G}(\omega,p)-{\cal
G}(\omega+\Omega,p+q)].$ The latter identity ensures that all
higher order symmetrized polarization bubbles for the
one-dimensional electron gas vanish thus turning the RPA
approximation into an exact procedure \cite{DL}. Analogously, the
identity (\ref{prop0}) helps to simplify the integrals
(\ref{JN11}).

In what follows we will employ the procedure similar to that applied
within the imaginary time formalism \cite{Kopietz}. Let us interchange
$\{\omega_1,\dots,\omega_N\}$ $N-1$ times in the following way. We first
put $\omega_N$ in front of $\omega_1$, then we place it in-between
$\omega_1$ and $\omega_2,$  and so on. This set of changes is summarized
in the table:
\begin{eqnarray}
\begin{array}{lllllll}
    &\{\omega_1 & \omega_2 &\omega_3 & \omega_4 & \dots & \omega_N\} \\
    &         &            &       &    \downarrow     &          &   \\
(1) & \{\omega_N & \omega_1 & \omega_2 & \omega_3 &  \dots &
\omega_{N-1}\} \\ (2) & \{\omega_1 & \omega_N & \omega_2 & \omega_3 &
\dots & \omega_{N-1}\} \\ (3) & \{\omega_1 & \omega_2 & \omega_N &
\omega_3 & \dots & \omega_{N-1}\} \\ \dots &&&&&& \\ (N-1) & \{\omega_1 &
\omega_2 & \dots  & \omega_{N-2}& \omega_N & \omega_{N-1}\}.
\end{array}
\label{change}
\end{eqnarray}
Having made these changes in the integral (\ref{JN11}) we express
$J_N$ as a sum of  $N-1$ corresponding integrals divided by $N-1.$
Afterwards we apply the identity (\ref{prop0}) in each of these
integrals excluding $\omega_N$ from the arguments of the Fermi
functions. Specifically, in the term corresponding to the sequence
(1) in the table (\ref{change}) we split the product of the first
two Fermi functions $n(E)n(E+\omega_N)$, in the term (2) we split
the product of the second and the third Fermi functions, and so on.
Then we get
\begin{eqnarray}
&& J_N=\frac{1}{N-1}\int dx\int_{-E_{\rm min}}^\infty
\frac{dE}{2\pi}\int\frac{d\omega_1\dots d\omega_{N}}{(2\pi)^{N}}
\nonumber\\ &&
 \times\, [\tilde
f(-\omega_N)n(E)+\tilde f(\omega_N)n(E+\omega_N)] \nonumber\\ &&
\times\, n(E+\omega_N+\omega_1)\dots
n(E+\omega_N+\omega_1+\dots+\omega_{N-2}) \nonumber\\ && \times\,
\tilde a(\omega_1)\tilde a(\omega_2)\dots \tilde
a(\omega_{N})\,{\rm e}^{i(\omega_1+\dots+\omega_N)x} \nonumber\\
&& +\,\frac{1}{N-1}\sum_{k=2}^{N-1}\int dx\int_{-E_{\rm
min}}^\infty\frac{dE}{2\pi}\int\frac{d\omega_1\dots
d\omega_{N}}{(2\pi)^{N}} \nonumber\\ &&
n(E)n(E+\omega_1)\dots\times [\tilde
f(-\omega_N)n(E+\omega_1+\dots+\omega_{k-1}) \nonumber\\ && \tilde
f(\omega_N)n(E+\omega_1+\dots+\omega_{k-1}+\omega_N)] \nonumber\\
&& \times\,
n(E+\omega_1+\dots+\omega_{k-1}+\omega_N+\omega_k)\times\dots
\nonumber\\ && \times\,
n(E+\omega_1+\dots+\omega_{k-1}+\omega_N+\omega_k+\dots+\omega_{N-2})
\nonumber\\ && \times\, \tilde a(\omega_1)\tilde a(\omega_2)\dots
\tilde a(\omega_{N})\,{\rm
e}^{i(\omega_1+\dots+\omega_N)x},\hspace{0.5cm}
\end{eqnarray}
where we have defined $\tilde
f(\omega)=\frac{1}{2}\left(1+\coth\frac{\omega}{2T}\right)$.

Now let us perform the frequency shifts $\omega_j\to
\omega_j+\omega_N$ in all the terms as well as the energy shift
$E\to E+\omega_N$ in the second term of the first integral in
order to eliminate $\omega_N$ from the arguments of the Fermi
functions. As a result we arrive at the following expression
\begin{eqnarray}
&& J_N=\frac{1}{N-1}\int dx\int_{-E_{\rm
min}}^\infty\frac{dE}{2\pi}\int\frac{d\omega_1\dots
d\omega_{N}}{(2\pi)^{N}}
\nonumber\\ &&
\times\, [\tilde f(-\omega_N)\tilde
a(\omega_1-\omega_N)+\tilde f(\omega_N)\tilde a(\omega_1){\rm
e}^{i\omega_Nx}]
\nonumber\\ &&
\times\, n(E)n(E+\omega_1)\dots
n(E+\omega_1+\dots+\omega_{N-2})
\nonumber\\ &&
\times\, \tilde
a(\omega_2)\dots \tilde a(\omega_{N})\,{\rm
e}^{i(\omega_1+\dots+\omega_{N-1})x}
\nonumber\\ &&
+\,\frac{1}{N-1}\sum_{k=2}^{N-1}\int dx\int_{-E_{\rm
min}}^\infty\frac{dE}{2\pi}\int\frac{d\omega_1\dots
d\omega_{N}}{(2\pi)^{N}}
\nonumber\\ &&
\times\,[\tilde f(-\omega_N)\tilde
a(\omega_{k-1})\tilde a(\omega_k-\omega_N)
\nonumber\\ &&
 +\,\tilde
f(\omega_N)\tilde a(\omega_{k-1}-\omega_N)\tilde a(\omega_k)]
\nonumber\\ &&
n(E)n(E+\omega_1)\dots n(E+\omega_1+\dots+\omega_{N-2})
\nonumber\\ &&
\times\, \tilde a(\omega_1)\tilde a(\omega_2)\dots\tilde
a(\omega_{k-2})\tilde a(\omega_{k+1})\dots \tilde a(\omega_{N})
\nonumber\\ &&
\times\, \,{\rm e}^{i(\omega_1+\dots+\omega_{N-1})x}
\nonumber\\ &&
+\,\frac{1}{N-1}\int dx\int_{-E_{\rm min}+\omega_N}^{-E_{\rm
min}}\frac{dE}{2\pi}\int\frac{d\omega_1\dots d\omega_{N}}{(2\pi)^{N}}
\nonumber\\ &&
\times\, n(E)n(E+\omega_1)\dots
n(E+\omega_1+\dots+\omega_{N-2})
\nonumber\\ &&
\times\, \tilde
f(\omega_N)\tilde a(\omega_1)\tilde a(\omega_2)\dots \tilde
a(\omega_{N})\,{\rm e}^{i(\omega_1+\dots+\omega_{N})x}.
\end{eqnarray}
We will assume that $|\omega_1|,\dots,|\omega_N|\ll E_{\rm min}$
which implies that $n(E)n(E+\omega_1)\dots
n(E+\omega_1+\dots+\omega_{N-2})=1$ in the interval $-E_{\rm
min}<E<-E_{\rm min}+\omega_N.$ This fact allows us to evaluate the
integral over $E$ in the last term. In addition, one observes
that $\tilde f(\omega_N)+\tilde f(-\omega_N)=1$ and, hence, the
function $\tilde f_n$ should drop out from all the other terms. We
obtain
\begin{eqnarray}
&& J_N=\frac{1}{N-1}\int dx\int_{-E_{\rm
min}}^\infty\frac{dE}{2\pi}\int\frac{d\omega_1\dots
d\omega_{N}}{(2\pi)^{N}}
\nonumber\\ &&
\times\, \tilde a(\omega_1)\dots
\tilde a(\omega_{N})\,{\rm e}^{i(\omega_1+\dots+\omega_{N-1})x}
\sum_{j=1}^{N-1} \frac{\tilde a(\omega_j-\omega_N)}{\tilde a(\omega_j)}
\nonumber\\ &&
\times\, n(E)n(E+\omega_1)\dots
n(E+\omega_1+\dots+\omega_{N-2})
\nonumber\\ &&
 -\,\frac{1}{N-1}\int
dx\int\frac{d\omega_1\dots d\omega_{N}}{(2\pi)^{N+1}}\,\omega_N\tilde
f(\omega_N)
\nonumber\\ &&
\times\, \tilde a(\omega_1)\tilde a(\omega_2)\dots
\tilde a(\omega_{N})\,{\rm e}^{i(\omega_1+\dots+\omega_{N})x}.
\end{eqnarray}

Returning  to the coordinate representation and assuming $a(x)\to
0$ for $|x|\to\infty$, we find
\begin{eqnarray}
&& J_N=\frac{1}{N-1}\int dx_1\dots dx_{N-1}\, \rho_0(x_1-x_2) \nonumber\\
&& \times\, \rho_0(x_2-x_3)\dots
\rho_0(x_{N-2}-x_{N-1})\rho_0(x_{N-1}-x_1) \nonumber\\ && \times\,
\prod_{j=1}^{N-1}a(x_j)\left(\sum_{j=1}^{N-1}a(x_j)\right) \label{JN1}\\
&& -\,\int
\frac{dxdy}{2\pi}\,\alpha(x-y)\frac{a^{N-1}(x)a(y)+a(x)a^{N-1}(y)}{4(N-1)}.
\hspace{0.5cm}\nonumber
\end{eqnarray}

To summarize, making use of Eq. (\ref{prop0}) we have reduced the
number of the functions $\rho_0$ under the integral by one. One
can employ this procedure further and, again applying Eq.
(\ref{prop0}), finally reduce the number of the functions $\rho_0$
down to two. As a result, $J_N$ takes the form
\begin{eqnarray}
J_N=\int dx
dy\rho_0(x-y)\rho_0(y-x)\sum_{k=1}^{N-1}\gamma_{kN}a^k(x)a^{N-k}(y)
\nonumber\\ +\int\frac{dxdy}{2\pi}\alpha(x-y)\sum_{k=1}^{N-1}
\beta_{kN}a^k(x)a^{N-k}(y).\hspace{0.5cm} \label{JN5}
\end{eqnarray}

Now let us apply the identity
\begin{equation}
\rho_0(x-y)\rho_0(y-x)=\rho_0(0)\delta(x-y)-\frac{\alpha(x-y)}{4\pi},
\label{rhorho}
\end{equation}
which  follows directly from Eq. (\ref{prop0}), and regroup the
terms in Eq. (\ref{JN5}).  Then we obtain
\begin{eqnarray}
&& J_N=\int dx \rho_0(0)u_Na^N(x) \nonumber\\ &&
+\,\int\frac{dxdy}{2\pi}\,\alpha(x-y)\sum_{k=1}^{N-1}\alpha_{kN}a^k(x)a^{N-k}(y).
\hspace{0.5cm} \label{JN2}
\end{eqnarray}

In order to find the coefficients $u_N$ and $\alpha_{kN}$ let us
rewrite Eq. (\ref{JN1}) in the form
\begin{eqnarray}
&& J_N[a(x)]=\left.-i\frac{d}{d\lambda}\frac{J_{N-1}[a(x){\rm e}^{i\lambda
a(x)}]}{N-1}\right|_{\lambda=0}
\nonumber\\ &&
-\,\int
\frac{dxdy}{2\pi}\,\alpha(x-y)\frac{a^{N-1}(x)a(y)+a(x)a^{N-1}(y)}{4(N-1)},
\nonumber
\end{eqnarray}
and derive the following equations:
\begin{eqnarray}
&& \alpha_{1,N}=\frac{N-2}{N-1}\alpha_{1,N-1}-\frac{1}{4(N-1)},
\nonumber\\ &&
\alpha_{N-1,N}=\frac{N-2}{N-1}\alpha_{N-2,N-1}-\frac{1}{4(N-1)},
\nonumber\\ &&
\alpha_{k,N}=\frac{k-1}{N-1}\alpha_{k-1,N-1}-\frac{N-1-k}{N-1}\alpha_{k,N-1},
\nonumber\\ && u_N=u_{N-1}
\end{eqnarray}
together with the boundary conditions $u_1=1$ and
$\alpha_{1,2}=-1/2.$ These equations can be solved with the result
\begin{equation}
u_n=1,\;\;\alpha_{kN}=-\frac{N}{4k(N-k)}.
\end{equation}
Let us now recall that the function $\rho_0(y-x)$ in Eq. (\ref{S0}) is
multiplied by ${\rm e}^{i\frac{\varphi^+(y)-\varphi^+(x)}{2}}.$ Hence, in
Eq. (\ref{JN2}) $\rho_0(0)$ has to be replaced by $$ \lim_{x\to
y}\rho_0(y-x)e^{i\frac{\varphi^+(y)-\varphi^+(x)}{2}}=\rho_0(0)+
\frac{\dot\varphi^+}{4\pi}. $$ This observation brings us to the following
expression for $J_N$:
\begin{eqnarray}
&& J_N=\int dx\,\left(\rho_0(0)+\frac{\dot\varphi^+(x)}{4\pi}\right)a^N(x)
\nonumber\\ &&
-\,\int dxdy\,
\alpha(x-y)\sum_{k=1}^{N-1}\frac{Na^k(x)a^{N-k}(y)}{8\pi k(N-k)}. \label{JN3}
\end{eqnarray}
Substituting $J_N$ into Eq. (\ref{S01}), we finally arrive at Eq.
(\ref{s0}).

\section{Weak reflection limit}

Let us derive the effective action of a coherent scatterer in the
weak reflection limit making use of the perturbation theory in
$R_n.$ In order to proceed we define the operator
\begin{eqnarray}
D_a(x,y)= \langle x|\left[1+\hat\rho_0\hat
a\right]^{-1}\hat\rho_0|y\rangle,  \label{D0}
\end{eqnarray}
where  $\langle x|\hat\rho_0|y\rangle=\rho_0(x-y)$ and $\langle
x|\hat\rho\hat a|y\rangle=\rho_0(x-y)a(y).$ Expanding Eq.
(\ref{S1}) in $\hat r^{\prime\dagger}\hat r'$ one finds
\begin{eqnarray}
&& iS_{\rm sc}^{(1)}=-2{\cal R}\int dx
\nonumber\\ &&\times\,
 \bigg[\lim_{x\to
y}\left(D_{a_1}(x,y){\rm
e}^{-i\frac{\varphi^+(x)-\varphi^+(y)}{2}}\right)a_1(x)
\nonumber\\ &&
+\,\lim_{x\to y}\left(D_{a_2}(x,y){\rm
e}^{i\frac{\varphi^+(x)-\varphi^+(y)}{2}}\right)a_2(x)\bigg]
\nonumber\\ &&
+\, 8{\cal R}\int dxdy\, D_{a_2}(x,y)D_{a_1}(y,x)\,{\rm
e}^{i[\varphi^+(x)-\varphi^+(y)]}
\nonumber\\ && \times\,
\sin\frac{\varphi^-(x)}{2}\sin\frac{\varphi^-(y)}{2}.\hspace{0.5cm}
\label{Ssc2}
\end{eqnarray}
Here we have defined $a_{1,2}(x)=\theta(t-x)\theta(x)\big({\rm e}^{\pm
i\varphi^-(x)}-1\big),$  and $D_a(x,y)=\langle x | \hat D_a | y\rangle$.

The function $D_a(x,y)$ can be found in the following way.
According to Eq. (\ref{S0}) one has
\begin{eqnarray}
iS_0[a+\delta a,0]=iS_0[a,0]+2\int dx\, D(x,x)\delta a(x) \nonumber\\
-\int dxdy\, D(x,y)\delta a(y)D(y,x)\delta a(x)+{\cal O}(\delta a^3(x)).
\label{S0da1}
\end{eqnarray}
On the other hand, from Eq. (\ref{s0}) we find
\begin{eqnarray}
&& iS_0[a+\delta a,0]=iS_0[a,0] -\int dx\frac{\rho_0(0)}{(1+
a(x))^2}\delta a^2(x) \nonumber\\ && +\,\int
dx\frac{2\rho_0(0)+\frac{1}{2\pi}\int dy
\alpha(x-y)\ln[1+a(y)]}{1+a(x)}\delta a(x) \nonumber\\ &&
+\,\int\frac{dxdy}{4\pi}\alpha(x-y)\bigg[\frac{\delta a(x)\delta
a(y)}{(1+a(x))(1+a(y))} \nonumber\\ && -\,\frac{\delta
a^2(x)\ln[1+a(y)]}{2(1+a(x))^2}-\frac{\delta
a^2(y)\ln[1+a(x)]}{2(1+a(y))^2}\bigg] \nonumber\\ && +\,{\cal O}(\delta
a^3(x)). \label{S0da2}
\end{eqnarray}
Comparing Eqs. (\ref{S0da1}) and (\ref{S0da2}) we obtain
\begin{eqnarray}
D_a(x,x)=\frac{\rho_0(0)+\frac{1}{4\pi}\int dy \alpha(x-y)\ln[1+a(y)]}{1+a(x)},
\label{Dxx}
\end{eqnarray}
\begin{eqnarray}
&& D_a(x,y)D_a(y,x)=
\nonumber\\ &&
=\bigg[\rho_0(0)+\int\frac{dy}{4\pi}\alpha(x-y)\ln[1+a(y)]\bigg]
\frac{\delta(x-y)}{(1+a(x))^2}
\nonumber\\ &&
-\,\frac{\alpha(x-y)}{4\pi(1+a(x))(1+a(y))}.
\label{DD}
\end{eqnarray}
These relations are consistent with the following form
of $D_a(x,y)$:
\begin{eqnarray}
D_a(x,y)&=& \frac{{\rm e}^{\frac{iT}{2}\int dz\big(\coth\pi
T(x-z)-\coth\pi T(y-z)\big)\ln[1+a(z)]}} {\sqrt{(1+a(x))(1+a(y))}}
\nonumber\\&& \times\, \rho_0(x-y)\,.
\label{Da}
\end{eqnarray}
Since Eqs. (\ref{Dxx},\ref{DD}) do not uniquely determine the function
$D_a(x,y)$, Eq. (\ref{Da}) has to be additionally checked. To this end we
make a shift $a(x)\to a(x)+\delta a(x)$ and find the linear in $\delta
a(x)$ correction  to $D_a(x,y)$. From Eq. (\ref{D0}) we obtain
\begin{eqnarray}
\delta D_a(x,y)=-\int dz\,D_a(x,z)\delta a(z)D_a(z,y).
\label{da}
\end{eqnarray}
Defining the function $f(x)=\frac{1}{2}\delta(x)-\frac{i}{2}T\coth\pi Tx,$
and making use of the identity
\begin{equation}
\rho_0(x-y)\rho_0(y-z)=[f(x-y)+f(y-z)]\rho_0(x-z),
\end{equation}
which is a direct consequence of Eq. (\ref{prop0}), one can verify
that the property (\ref{da}) also holds for the function
(\ref{Da}). We then conclude that the two functions (\ref{D0}) and
(\ref{Da}) may differ only by a shift of the argument $a(x)$. The
latter shift is zero since in both cases at $a(x)=0$ one gets
$D_{a=0}(x,y)=\rho_0(x-y).$ Thus, we conclude that Eq. (\ref{Da})
is indeed correct.

What remains is to substitute this result into Eq. (\ref{Ssc2})
and get
\begin{eqnarray}
iS_{\rm sc}^{(1)}&=&-4{\cal R}\int_0^t dx\rho_0(0)[1-\cos\varphi^-(x)]
\nonumber\\ && +\,\frac{i{\cal R}}{\pi}\int_0^t
dx\dot\varphi^+(x)\sin\varphi^-(x) \nonumber\\ && +\frac{\cal
R}{\pi}\int_0^t dxdy\alpha(x-y)\varphi^-(x)\sin\varphi^-(y) \nonumber\\ &&
+\,8{\cal R}\int_0^t dxdy\rho_0(x-y)\rho_0(y-x){\rm
e}^{i[\varphi^+(x)-\varphi^+(y)]} \nonumber\\ &&\times\, {\rm e}^{\int_0^t
dz[T\coth\pi T(x-z)-T\coth\pi T(y-z)]\varphi^-(z)} \nonumber\\ &&\times\,
\sin\frac{\varphi^-(x)}{2}\sin\frac{\varphi^-(y)}{2}.
\end{eqnarray}
In the last term of this equation we again apply the identity
(\ref{rhorho}) and arrive at the final expression for the
effective action (\ref{action}).

\end{document}